\documentclass[twocolumn,superscriptaddress,nofootinbib,aps,floatfix,preprintnumbers,amsmath,amssymb]{revtex4-1}

\usepackage{slashed}
\usepackage{latexsym}
\usepackage{graphics}
\usepackage{bm}
\usepackage{graphicx} 
\usepackage{color}
\usepackage{amsmath}
\usepackage{amssymb}
\usepackage{feynmp-auto}
\usepackage{slashed}
\usepackage[dvipsnames]{xcolor}
\usepackage{xcolor}
\graphicspath{{figs/}}
\usepackage{enumitem}
\usepackage{braket}
\usepackage{ulem}
\usepackage{hyperref}

\usepackage{amsfonts, amssymb,amsmath,latexsym,graphics, graphicx,epsfig,multirow,comment,
,feyn,slashed,xcolor,afterpage, makecell} 
\usepackage{booktabs, blindtext}
\usepackage{tabularx,afterpage}
\usepackage{enumerate}
\newcommand{\beq}{\begin{equation}}
\newcommand{\eeq}{\end{equation}}
\newcommand{\bea}{\begin{eqnarray}}
\newcommand{\eea}{\end{eqnarray}}

\newcommand{\Fmolsq}{\ifmmode {\lvert F_{\rm mol}(q) \rvert}^2 \else ${\lvert F_{\rm mol}(q) \rvert}^2$\fi}
\newcommand{\FmolvJsq}{\ifmmode {\lvert F_{{\rm mol}, v' J'}(q) \rvert^2} \else ${\lvert F_{{\rm mol}, v' J'}(q) \rvert^2}$\fi}
\newcommand{\FmolvJsqAvg}{\ifmmode {\langle\lvert F_{{\rm mol}, v' J'}(q,T) \rvert^2\rangle} \else ${\langle\lvert F_{{\rm mol}, v' J'}(q,T) \rvert^2\rangle}$\fi}

\begin{document}
 
\title{Dark Matter Direct Detection with Quantum Dots}

\author{Carlos Blanco}
\affiliation{Department of Physics, Princeton University, Princeton, NJ 08544, USA}
\affiliation{Stockholm University and The Oskar Klein Centre for Cosmoparticle Physics,  Alba Nova, 10691 Stockholm, Sweden}

\author{Rouven Essig}
\affiliation{C.~N.~Yang Institute for Theoretical Physics, Stony Brook University, USA}

\author{Marivi Fernandez-Serra}
\affiliation{Department of Physics and Astronomy, Stony Brook University, Stony Brook, NY 11794-3800, USA}
\affiliation{Institute for Advanced Computational Science, Stony Brook University, Stony Brook, NY 11794-3800, USA}

\author{Harikrishnan Ramani}
\affiliation{Stanford Institute for Theoretical Physics, Department of Physics, Stanford University, Stanford, CA 94305, USA}

\author{Oren Slone}
\affiliation{Department of Physics, Princeton University, Princeton, NJ 08544, USA}
\affiliation{Center for Cosmology and Particle Physics, Department of Physics, New York University, New York, NY 10003, USA}

\date{\today}

\begin{abstract}
We propose using Quantum Dots as novel targets to probe sub-GeV dark matter-electron interactions.  Quantum dots are nanocrystals of semiconducting material, which are commercially available, with gram-scale quantities suspended in liter-scale volumes of solvent. Quantum dots can be efficient scintillators, with near unity single-photon quantum yields, and their band-edge electronic properties are determined by their characteristic size, which can be precisely tuned. Examples include lead sulfide (PbS) and lead selenide (PbSe) quantum dots, which can be tuned to have sub-eV optical gaps. A dark-matter interaction can generate one or more electron-hole pairs (excitons), with the multi-exciton state decaying via the emission of two photons with an efficiency of about 10\% of the single-photon quantum yield.  An experimental setup using commercially available quantum dots and two photo-multiplier-tubes (PMTs) for detecting the coincident two-photon signal can already improve on existing dark-matter bounds, while using photodetectors with lower dark-count rates can improve on current constraints by orders of magnitude. 

\end{abstract}

\maketitle

\section{Introduction}
\label{sec:intro}

Experimental direct-detection concepts for Dark Matter (DM) below the GeV scale have been extensively studied in recent years (see, e.g.,~\cite{Essig:2011nj,Essig:2012yx,An:2014twa,Lee:2015qva,Essig:2015cda,Hochberg:2015pha,Hochberg:2015fth,Derenzo:2016fse,Bloch:2016sjj,Hochberg:2016ntt,Hochberg:2016ajh,Hochberg:2016sqx,Kahn:2016aff,Tiffenberg:2017aac,Knapen:2017ekk,Hochberg:2017wce,Knapen:2017xzo,Crisler:2018gci, Agnes:2018oej,Bringmann:2018cvk,Griffin:2018bjn,Essig:2019xkx,Abramoff:2019dfb,Emken:2019tni,Trickle:2019ovy,Griffin:2019mvc,Trickle:2019nya,Coskuner:2019odd,SENSEI:2020dpa,Bernstein:2020cpc,Du:2020ldo,Mitridate:2021ctr,Griffin:2021znd,Coskuner:2021qxo,Berghaus:2021wrp,Aguilar-Arevalo:2022kqd,An:2013yfc,Angle:2011th,Graham:2012su,Aprile:2014eoa,Aguilar-Arevalo:2016zop,Cavoto:2016lqo,Kouvaris:2016afs,Robinson:2016imi,Ibe:2017yqa,Dolan:2017xbu,Romani:2017iwi,Budnik:2017sbu,Bunting:2017net,Cavoto:2017otc,Fichet:2017bng,Emken:2017erx,Emken:2017hnp,Emken:2017qmp,Emken:2018run,LUX:2018akb,Agnese:2018col,Settimo:2018qcm,Ema:2018bih,Akerib:2018hck,CDEX:2019hzn,EDELWEISS:2019vjv,Bell:2019egg,Liu:2019kzq,XENON:2019zpr,Baxter:2019pnz,Aguilar-Arevalo:2019wdi,Armengaud:2019kfj,Aprile:2019jmx,Kurinsky:2019pgb,Cappiello:2019qsw,Catena:2019gfa,Lin:2019uvt,Blanco:2019lrf,Geilhufe:2019ndy,Griffin:2020lgd,Trickle:2020oki,Hochberg:2021pkt,Akerib:2021pfd,LUX:2020yym,Liang:2020ryg,GrillidiCortona:2020owp,Ma:2019lik,Nakamura:2020kex,Collar:2021fcl,Hochberg:2021ymx,Kahn:2021ttr,Blanco:2021hlm,SuperCDMS:2022kgp,Redondo:2013lna,Amaral:2020ryn,Arnaud:2020svb,Aprile:2016wwo,Aprile:2019xxb,XENON:2021myl,PandaX-II:2021nsg,DAMIC:2016qck, DAMIC:2019dcn,Knapen:2021bwg,Alexander:2016aln,Battaglieri:2017aum,BRNreport,BRNreportdetector,Essig:Physics2020,Essig:2022dfa}). To date, there are a number of leading search strategies, many of which focus on DM-electron scattering as an efficient method to extract and detect small energy deposits following either a scattering or absorption event with a DM particle. Since the ultimate sensitivity of each strategy is unknown, and since there are several unexplained low-energy backgrounds~\cite{excessworkshop} that may impact different DM detection techniques in distinct ways, it is imperative to have a multitude of approaches to search for sub-GeV DM. 
When designing an experiment it is beneficial to have a low detection threshold to probe low-mass DM, excellent control over backgrounds, and a target that can be scaled up in mass to probe smaller interactions. 

Scintillating targets have previously been argued to potentially fulfill these requirements~\cite{Derenzo:2016fse}. For example, semiconductor scintillators such as GaAs, have been proposed as a promising target due to their low band gap~\cite{Derenzo:2016fse,Derenzo:2018plr,Vasiukov:2019hwn,Derenzo:2020hfo,GaAsLOI,Derenzo:2022cnn}. Other examples proposed in the literature include directional targets such as organic crystals~\cite{blanco2020dark}, as well as targets which can probe DM-nucleon interactions such as inorganic compounds~\cite{Budnik:2017sbu,Essig:2016crl} and molecular gas targets~\cite{Arvanitaki:2017nhi,Essig:2019kfe}. Finally, molecular liquid scintillators have been experimentally demonstrated to be rapidly deployable and scalable, and have already been used to constrain sub-GeV DM~\cite{Blanco:2019lrf,Miramonti:2006kp}.  

This study advocates for \textit{quantum dots} (QDs), nanoscopic crystals of semiconducting material, as another excellent low-threshold scintillating DM detector target. 
QDs are commercially available, with gram-scale quantities suspended in liter-scale volumes of solvent. The basic concept is that DM-electron scattering or, for the case of bosonic DM, absorption  by an electron, excites an electron-hole pair in the QD which subsequently relaxes by emitting light. The optical properties of a QD are dependent on its size and shape, which can be tuned precisely at commercial scales.

QDs have similar optical and electronic properties to molecular scintillators. Specifically, their optical properties when suspended in a solvent and their photoluminescence Stokes shift (shift to lower energy of the emission spectrum compared to the absorbed light) make them very efficient emitters of photons that can easily escape from a bulk target. Essentially, their relaxation properties are very much akin to molecular fluorescence. Additionally, QDs benefit from the DM-scattering and absorption characteristics of bulk semiconductors; when the DM scattering takes place even modestly above threshold, the QD's electronic response is essentially that of a bulk semiconductor. Practically, an experiment using QDs as a scintillating target could immediately take advantage of the scalability of the target mass and the ease of detecting the photon signal (similar to molecular scintillators~\cite{Blanco:2019lrf,blanco2020dark}), while utilizing the small bandgap of  semiconductors.

Following an interaction with DM, an electron-hole pair (an exciton) is created. If this initial exciton has energy larger than twice the energy gap of the QD, it can relax by generating several more excitons with lower energies. This multi-exciton generation can yield multiple time-coincident photons as a detectable signal, which would have intrinsically low background. While this would be the most straightforward search strategy for a setup using currently existing technology, future ultra-low-background photodetectors capable of sensing single photons could allow one to search for single-photon fluorescence, which benefits from significantly higher quantum yields.

The paper is organized as follows. In Sec.~\ref{sec:QDFluor}, we describe the salient features of QDs in the context of direct-detection experiments. In Sec.~\ref{sec:QD_Signal_Rates}, we describe the calculations of the expected signal rates for both scattering and absorption events. Sec.~\ref{sec:BGs} describes some of the expected backgrounds rates for the proposed setup. Sec.~\ref{sec:sens} presents the results for the expected sensitivity to DM parameter space for the proposed setup. Finally, we conclude and discuss the main results in Sec.~\ref{sec:conclusions}. Details on a simple semi-analytic treatment of the DM-electron scattering rates in QDs as well as a detailed discussion of calculations used in our results are presented in the appendices.

\section{Direct Detection with Quantum Dots}
\label{sec:QDFluor}

\begin{figure}[t]
    \centering
    \includegraphics[width=\columnwidth]{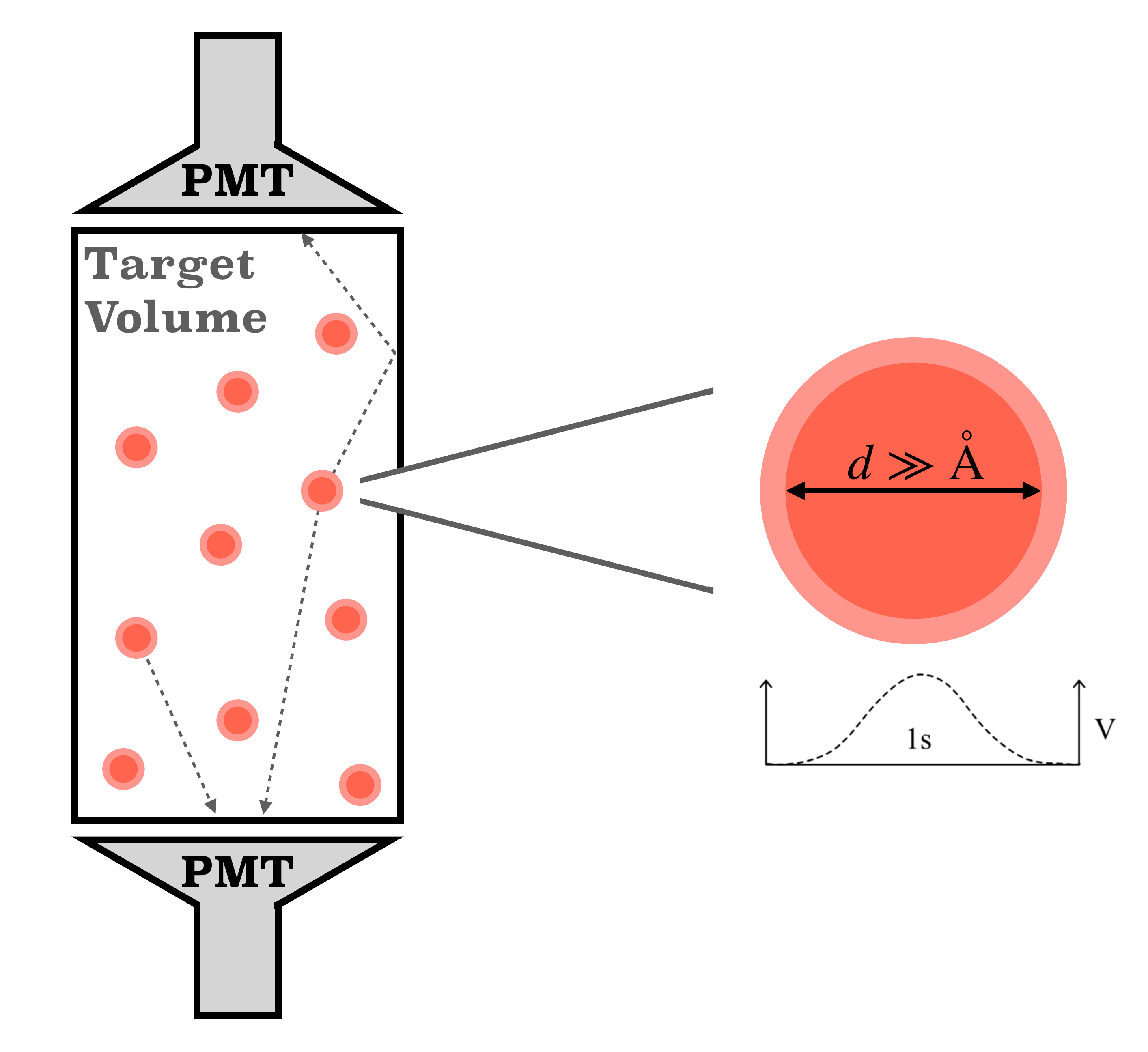}
    \caption{Envisioned experimental setup. Spherical quantum dots are in a colloidal suspension (suspended in solvent) at the maximal concentration such that the volume remains transparent to florescent photons emitted from the quantum dots. The bases of the detector are coupled to photodetectors, such as PMTs, while the remaining surface area is coated with reflective material. The quantum dots are synthesized with typical diameter, $d$, that is much larger than the size of an atom. Also shown is the approximate form of the 1s wavefunction of a spherical quantum dot.}
    \label{fig:QDdiagram}
\end{figure}

While various types of QDs can be fabricated, this study focuses on colloidal QDs composed of elements from columns IV and VI in the periodic table (also known as IV-VI QDs). In particular, for results presented in this study, we consider PbS QDs, which are commercially available, well-studied, and infrared-active (although there are many other types of QDs with similar properties, e.g., PbSe). Individual QDs of this type have near-unity single-photon quantum yield, and their colloidal suspensions can have bulk photoluminescent quantum yields as high as 0.5. These colloidal suspensions are synthesized from solutions of precursors and are available as stable suspensions of QDs in a solvent with a specific QD concentration and a specific crystal size~\cite{moreels2009size,hines2003colloidal}. The synthesis generally proceeds through the nucleation of monomeric precipitates of the QD material followed by a period of crystal growth in suspension. Techniques that control the crystal growth phase can generate suspensions of QDs that have very precise size characteristics. The result is a suspension of nanocrystals, each of which consists of hundreds to thousands of atoms.

All calculations and estimates in this study consider the simple case of spherical QDs, since these are commercially available and well-studied. However, QDs can also be synthesized in other morphologies which may have further interesting optical properties such as a higher quantum yield. In practice, the term ``spherical QDs" refers to objects that are only approximately spherical. In reality, the surface of any QD is covered with ligands, which serve to passivate the surface (allowing charges from surface-dangling bonds to be effectively neutralized), while also maintaining the QD in a colloidal suspension. Importantly, the optical emission characteristics of colloidal QDs is very sensitive to the surface properties of the crystal. For this reason and also because of the inherent complexity of QDs, many of the estimates presented in this study are based on measured quantities in the literature and not on first-principle calculations.

A DM scattering or absorption event with the electronic ground state of a colloidal QD results in an excitonic state that is typically well above the optical gap edge. As will be discussed below, the properties of such excitonic states are almost identical to those of a bulk semiconductor and therefore calculating their properties follows directly from well studied techniques used for semiconductors (see, e.g.,~Ref.~\cite{Essig:2015cda}).

A schematic diagram of the envisioned experimental setup is shown in Fig.~\ref{fig:QDdiagram}. The spherical QDs are in a colloidal suspension at a concentration for which the volume remains transparent to fluorescent photons emitted from the QDs. The tank is envisioned as a cylindrical volume whose bases are photon collection surfaces coupled to photodetectors such as PMTs, while the remaining surface area is coated with a highly reflective material such as titanium oxide or sodium silicate~\cite{Collar:2018ydf}. A DM particle entering the detector volume can interact, either via scattering or absorption, with an electron in one of the suspended QDs. The energy transferred to the QDs results in the creation of excitonic states.

\begin{figure}
    \centering
    \includegraphics[width=\columnwidth]{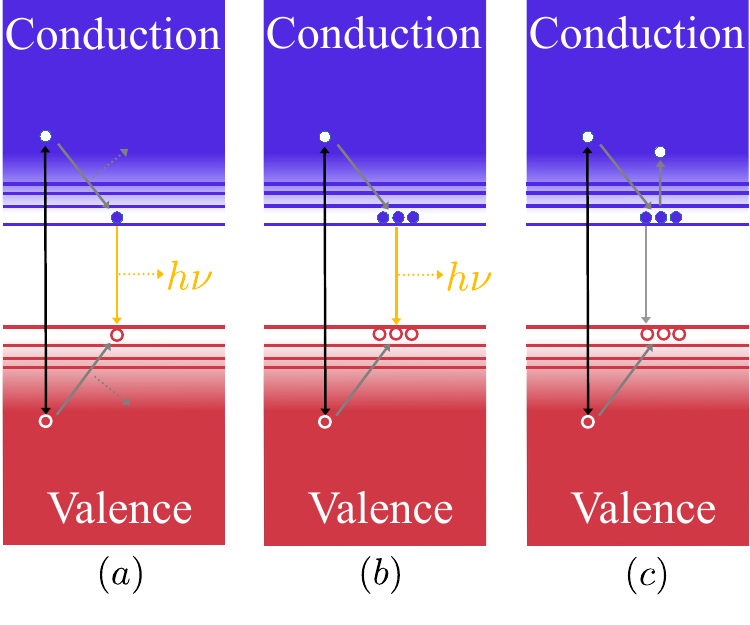}
    \caption{The radiative and non-radiative transition channels for QDs. (a) The excitation of a primary hot electron-hole pair (exciton), followed by cooling via non-radiative phonon emission, which produces a band-edge exciton. The exciton then spontaneously emits a photon (fluorescence) with quantum yield QY$_{\gamma} \approx 0.5$~\cite{yang2017iodide}. (b) The primary hot exciton undergoes multi-exciton generation (MEG), creating several electron-hole pairs. The multi-exciton state can spontaneously emit two coincident photons with quantum yield QY$_{\gamma\gamma} \approx 0.05$~\cite{beyler2014sample}. (c) One of the electron-hole pairs undergoes Auger recombination, exciting one of the other carriers and depleting the multi-exciton without emitting a photon. This process is efficient and is the main reason why QY$_{\gamma\gamma}$ is small in comparison to QY$_\gamma$.}
    \label{fig:transitions}
\end{figure}

After an energetic (hot) exciton has been generated, it can relax via a number of channels into a band-edge state. Possible exciton relaxation processes are shown in Fig.~\ref{fig:transitions}. Relaxation to the band-edge state can happen either through the emission of phonons, creating a single optically-gapped exciton at the band edge (panel (a)), or by generating multiple optically-gapped excitons at the band edge while also emitting phonons to balance the energy budget (panel (b)). The latter case is known as multi-exciton generation (MEG) and has been experimentally shown to be efficient in colloidal QDs~\cite{ellingson2005highly,schaller2004high,hardman2011electronic,midgett2013size,murphy2006pbte} (this is the analog of impact ionization in a bulk semiconductor). In either case, the band-edge excitons (either single or multiple) can radiatively recombine, producing fluorescence photons. In the event of MEG, there is a competing three-body relaxation process known as Auger recombination whereby an exciton recombines non-radiatively by donating its energy to a neighboring charge carrier (panel (c)). This process efficiently depletes the band-edge multi-exciton state~\cite{an2008excited}. However, since Auger recombination requires at least three carriers, the single band-edge exciton is safe from this non-radiative de-excitation. 

While MEG occurs in bulk semiconductors, strongly confined systems and in particular PbS QDs are expected to have a higher yield of excitons per deposited energy above the bandgap. Compared to the case of a bulk semiconductor, where there is more continuum overlap between the phonon states and electronic bands, in QDs the electronic band is broken into discrete states near the band edge, reducing the overlap and making phonon emission less favorable. Importantly, the existence of real intermediate states between primary excitation and final radiative emission allows one to separate the calculation of the DM-electron interaction rate from the calculation of the photon yield. Experimentally, it allows a separation of the bulk target from the photodetector.

The single-exciton radiative quantum yield, QY$_{\gamma}$, is an experimental measure of how many photons are produced per primary electron-hole pair. Since a single exciton can only radiatively de-excite through the emission of a single photon, this is an experimental measure of the radiative branching fraction of the single exciton. Similarly, the bi-exciton (two pairs of excitons) fluorescence quantum yield, QY$_{\gamma \gamma}$, is an experimental measure of how many coincident two-photon emissions are produced per bi-exciton. For the case of PbS, the value QY$_{\gamma}$ is measured to be QY$_{\gamma} \approx 0.5$~\cite{yang2017iodide}, while QY$_{\gamma\gamma}$ is estimated at QY$_{\gamma\gamma} \approx 0.05$~\cite{beyler2014sample}. As shown in the remainder of this paper, these values, together with exciton creation and photon collection efficiencies, are large enough to permit an experimental design based on currently available technology that could probe significant regions of unexplored DM parameter space.

One of the salient features of colloidal QDs is the Stokes shift between their excitation and emission spectra, which is similar to that of molecular chromophores (see, e.g.,~\cite{Greben2015}). For this reason, there is only a small overlap between the excitation and emission spectra, and fluorescent photons cannot efficiently create excitations in neighboring QDs. Experimentally, for PbS, this becomes evident in the bulk fluorescence quantum yield, as discussed below in Sec.~\ref{subsec:Photon_Collection} and App.~\ref{sec:QY}. Therefore, any sufficiently dilute bulk suspension of colloidal QDs is essentially transparent to its emission photons. This feature results in a fluorescent signal that is proportional to the target volume and allows for easy scalability of any experimental setup. The emitted photons can be detected by photodetectors on the surface of the target volume. 

The size-dependent optical properties of colloidal QDs makes their diameter, $d$, a parameter that can be used to optimize the detector's performance. Specifically, increasing $d$ decreases the confinement energy and therefore also the minimal energy required to excite the QD, $\Delta E_{\text{QD}}$. Additionally, increasing $d$ decreases the Stokes shift~\cite{Greben2015} and therefore also decreases the bulk fluorescence quantum yield. For the case of DM scattering, we choose $d=3.3\,\text{nm}$ PbS QDs, which have a measured bulk fluorescence quantum yield of about 20\%~\cite{Greben2015} and $\Delta E_{\text{QD}}$ approximately equal to the bandgap of silicon, one of our bulk semiconductor benchmarks. For the case of bosonic DM absorption, we choose $d=5\,\text{nm}$, since dielectric-function data exists for this size. Below, we present rate calculations and projected sensitivities for the envisioned setup.

\section{Quantum Dot Signal Rates}
\label{sec:QD_Signal_Rates}

\subsection{Electronic Structure}

Electrons in a periodic and infinite lattice occupy Bloch states given by 
\begin{equation}
    \Psi_{\text{bulk}}(\mathbf{r})=u_{\kappa}(\mathbf{r})e^{i\mathbf{k}\cdot \mathbf{r}}\,,
    \label{eq:Psi_bulk}
\end{equation}
where $u_{\kappa}(\mathbf{r})$ respects the translational symmetry of the lattice, and the exponential is an envelope solution satisfying the time-independent Schr\"odinger equation in vacuum. A QD is a nanocrystal whose characteristic diameter, $d$, is large compared to the lattice spacing. Therefore, locally, electrons experience a periodic potential. However, globally, the exact translational symmetry is broken by the finite size of $d$, and the envelope solution is no longer that given by Eq.~\eqref{eq:Psi_bulk}. Specifically, the long-wavelength behavior of the electronic system must obey new boundary conditions set by the surface of the QD.

Electrons in a QD occupy states in which the envelope function satisfies these new boundary conditions.  Assuming  a spherical QD, the electron wave function is approximately given by 
\begin{align}\label{eq:QDpsi1}
    \Psi_{\kappa,n\ell m}(\mathbf{r})=u_{\kappa}(\mathbf{r})\psi_{n\ell m}(\mathbf{r}),
\end{align}
where $\kappa \in \{c,v\}$ denotes either the conduction ($c$) or valence ($v$) bands. The function $\psi_{n\ell m}(\mathbf{r})$ is the envelope wavefunction,
\beq\label{eq:QDpsi2}
\psi_{n\ell m}(\mathbf{r})=\mathcal{R}_{n\ell}(r)\mathcal{Y}_{\ell m}\left(\theta,\phi \right)\,,
\eeq
where $\mathcal{Y}_{\ell m}\left(\theta,\phi \right)$ are spherical harmonics, and $\mathcal{R}_{n\ell}(r)$ are the radial wavefunctions for a free particle in a spherical well with infinite walls, 
\beq\label{eq:QDpsi3}
\mathcal{R}_{n\ell}(r)=\sqrt{\frac{2}{R^3}} \frac{j_\ell\left(\chi_{n\ell} r/R\right)}{j_{\ell+1} \left( \chi_{n\ell} \right)}\,.
\eeq
Here, $R=d/2$, $j_\ell$ are spherical Bessel functions, $\chi_{n\ell}$ is the $n^{\rm th}$ root of the equation $j_\ell(x)=0$, and $\{n,\ell,m\}$ are the quantum numbers of the envelope state. Fig.~\ref{fig:QDdiagram} shows a diagrammatic representation of the radial component of the 1s envelope wavefunction.

The description given in Eqs.~(\ref{eq:QDpsi1}--\ref{eq:QDpsi3}) is most accurate for the lowest lying states of the conduction and highest states of the valence band, since the long-wavelength solutions are those most affected by the boundary. This is known as quantum confinement. The energetic effect of quantum confinement is to effectively lift the band edge. In a bulk crystal with band gap $\Delta E_g$, the density of states scales with energy as $\rho_{\text{bulk}}(E)\propto\sqrt{E-\Delta E_g}$ above the band edge. In a QD, this vanishing continuum at the band edges becomes a discrete ladder of $\psi_{n\ell m}$ states. The energy gap between the highest valence QD state and the lowest conduction QD state (analogous to the ``highest occupied molecular orbital"-``lowest unoccupied molecular orbital" gap of molecules) is given by,
\begin{align}\label{eq:EgapQD}
    \Delta E_{\rm QD} \approx \Delta E_g + \frac{\pi^2}{2 R^2}\left(\frac{1}{m_{h,{\rm eff}}} + \frac{1}{m_{e,{\rm eff}}} \right),
\end{align}
where the second term is the energy of confinement of the electron-hole state with effective masses $m_{e,{\rm eff}}$ and $m_{h,{\rm eff}}$, respectively.

\begin{figure}
    \centering
    \includegraphics[width=0.725\columnwidth]{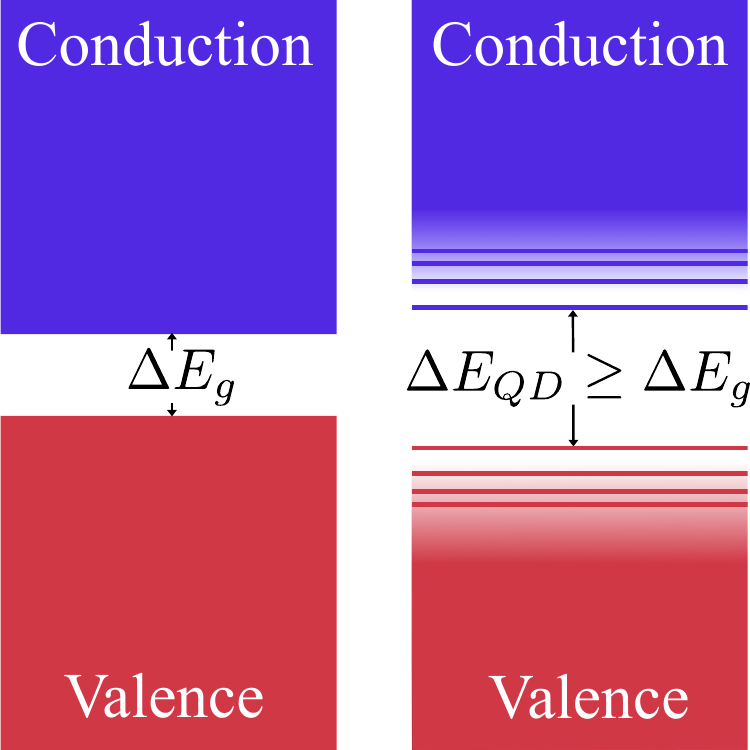}
    \caption{The effect of strong quantum confinement, for example in a QD, is shown diagrammatically. The left panel shows the continuum bands of an infinite crystal, while the right panel shows the effects of strong confinement of electrons in a finite crystal, which cause discrete levels to replace the continuum states at the band edges.}
    \label{fig:bands}
\end{figure}

Fig.~\ref{fig:bands} shows the effects of strong quantum confinement of the continuum bands of a semiconductor. This strong confinement is achieved when the size of the QD is much smaller than the Bohr radius of the exciton in the material. In other words, the size of the QD is smaller than the characteristic size of the bound electron-hole pair. For example, in PbS (PbSe), the size of the exciton is $20$~nm ($46$~nm)~\cite{stouwdam2007photostability}. In the strong-confinement regime, the electron and hole behave like uncorrelated particles.

Far away from the band edge, the properties of a QD converge to those of bulk semiconductors. Specifically, it has been shown experimentally that the optoelectronic response of PbS and PbSe QDs resemble those of bulk semiconductors for excitations above $E_r \gtrsim 2.5~\text{eV}$~\cite{moreels2007composition,moreels2007composition,grinbom2010density}. The same can be shown from a  numerical estimate, since the discrete QD density of states becomes dense and converges onto the bulk density of states even moderately above the band edges once any nanocluster grows beyond about 100~atoms~\cite{kane1996theoretical,debellis2017quantum} (a PbS QD with a diameter of $d=3$~nm contains about 1000 atoms). Heuristically, this means that short-wavelength states do not depend strongly on the boundary. Therefore, in practice the approximations of  Eqs.~(\ref{eq:QDpsi1}--\ref{eq:QDpsi3}) are not necessary for computing rates for almost all cases of interest in the current proposal. Instead, results are taken from more precise numerical codes designed to predict excitation rates in bulk semiconductors.

For completeness, App.~\ref{sec:semianalmod} provides some additional intuition regarding this point. A semi-analytic model of a QD as a band-edge dipole correction to the bulk semiconductor is described. It is expected that the leading correction to the bulk description of the form factor will be this dipole term. The appendix shows that changes to the dipole behavior of the form factor for an excitation driven by DM do not significantly affect the scattering rate, since that region is kinematically forbidden anyway. Namely, we show that in the limit when $q \gg 1/a \approx 2.5 \;\text{keV}$, both the bulk and QD have form factors that are dominated by momentum transfers localized to single atoms. Therefore, the rate is essentially independent of the envelope piece of the wave function. Furthermore, it is expected that when $q \gg 1/R \approx 0.5 \;\text{keV}$ and $E_r> 2.5\;\text{eV}$, the electronic response of the QD is that of the equivalent bulk semiconductor since the envelope piece of the wavefunction is a result of the boundary conditions at $R$.

\subsection{Scattering Rates}
\label{subsec:Rates}

The total directionally-averaged differential electronic excitation rate is given by,
\beq
\label{eq:diffRate}
    \frac{dR_{\rm ex}}{d\ln{E_r}} = \frac{N_t \rho_{\chi}\overline{\sigma}_e} {8m_{\chi} \mu_{e\chi}} \int_{q_{\rm min}}^{q_{\rm max}} q dq \eta(v_{\rm min}) |F_{\rm DM}(q)|^2  |f_{\rm QD}(q,E_r)|^2,  
\eeq
where $N_t$ is the number of target unit cells (defined in terms of lattice sites as opposed to number of electrons because of the normalization of $f_{\rm QD}$, see discussion below and Ref.~\cite{Essig:2015cda}), $\rho_\chi=0.3 \,\text{GeV}\,\text{cm}^{-3}$ is the local DM density, $\mu_{e\chi}$ is the reduced mass of the electron-DM system, $q$ is the momentum transfer to the QD, and $E_r$ is the recoil energy. The value of $N_t$ is related to the density of the target material (the mass per unit volume of QD material), $\rho_{\rm QD}$, and the target volume, $V$, by 
\beq
N_t = \rho_{\rm QD} \frac{V}{N_A (239.3\,\text{g}\,\text{mol}^{-1})}\,.
\label{eq:N_t}
\eeq

The normalized cross section, $\overline{\sigma}_e$, and DM form-factor, $F_{\rm DM}(q)$, are related to the free-electron DM scattering matrix element by~\cite{Essig:2011nj}
\begin{align}
    \overline{\left| \mathcal{M}_{\text{free}}(\mathbf{q}) \right|^2}&\equiv \overline{\left| \mathcal{M}_{\text{free}}(\alpha m_e) \right|^2} \times \left| F_{\rm DM}(q)\right|^2, \nonumber \\
    \overline{\sigma}_e &\equiv \frac{\mu_{e\chi}^2 \overline{\left| \mathcal{M}_{\text{free}}(\alpha m_e) \right|^2} }{16\pi m_{\chi}^2 m_{e}^2}\,,
\end{align}
where $m_\chi$, $m_e$ are the DM and electron masses, respectively, $\alpha$ is the fine structure constant, and $\alpha m_e$ is the typical momentum transfer for creation of an exciton in a QD. For the case of a DM particle that interacts with electrons through a massive mediator whose mass exceeds that of the DM particle, $F_{\rm DM} = 1$. For the case of a light mediator whose mass is much smaller than the momentum transfer from the DM to the electron, $F_{\rm DM} = (\alpha m_e / q)^2$. These two limits for $F_{\rm DM}$ will be used for all results presented below.

The function $\eta(v_{\rm min}(q))$ is the integrated velocity distribution defined as in~\cite{Essig:2015cda}, and $v_{\rm min}(q)$ is the $q$-dependent minimal DM velocity allowed by 
\beq
v_\text{min}(q,E_r) = \frac{E_r}{q} + \frac{q}{2 m_\chi}\,.
\eeq
We evaluate rates assuming a Maxwellian velocity distribution with values for the velocity of Earth, $v_{\rm Earth} = v_0+v_{\odot} = 251\text{ km/s}$, local mean speed $v_0 = 238\text{ km/s}$, escape velocity $v_\text{\rm esc} = 544\text{ km/s}$, and solar peculiar velocity $v_{\odot} \approx 13\text{ km/s}$~\cite{Baxter:2021pqo}. The $q$-integral in Eq.~(\ref{eq:diffRate}) is calculated over the kinematically allowed region enclosed by $q_{\rm min}\leq q \leq q_{\rm max}$, where the minimum and maximum values are given by the roots of $v_{\rm min}(q)=v_{\rm esc}+v_{\rm Earth}$.

\begin{figure*}
    \centering
    \includegraphics[width=0.45\textwidth]{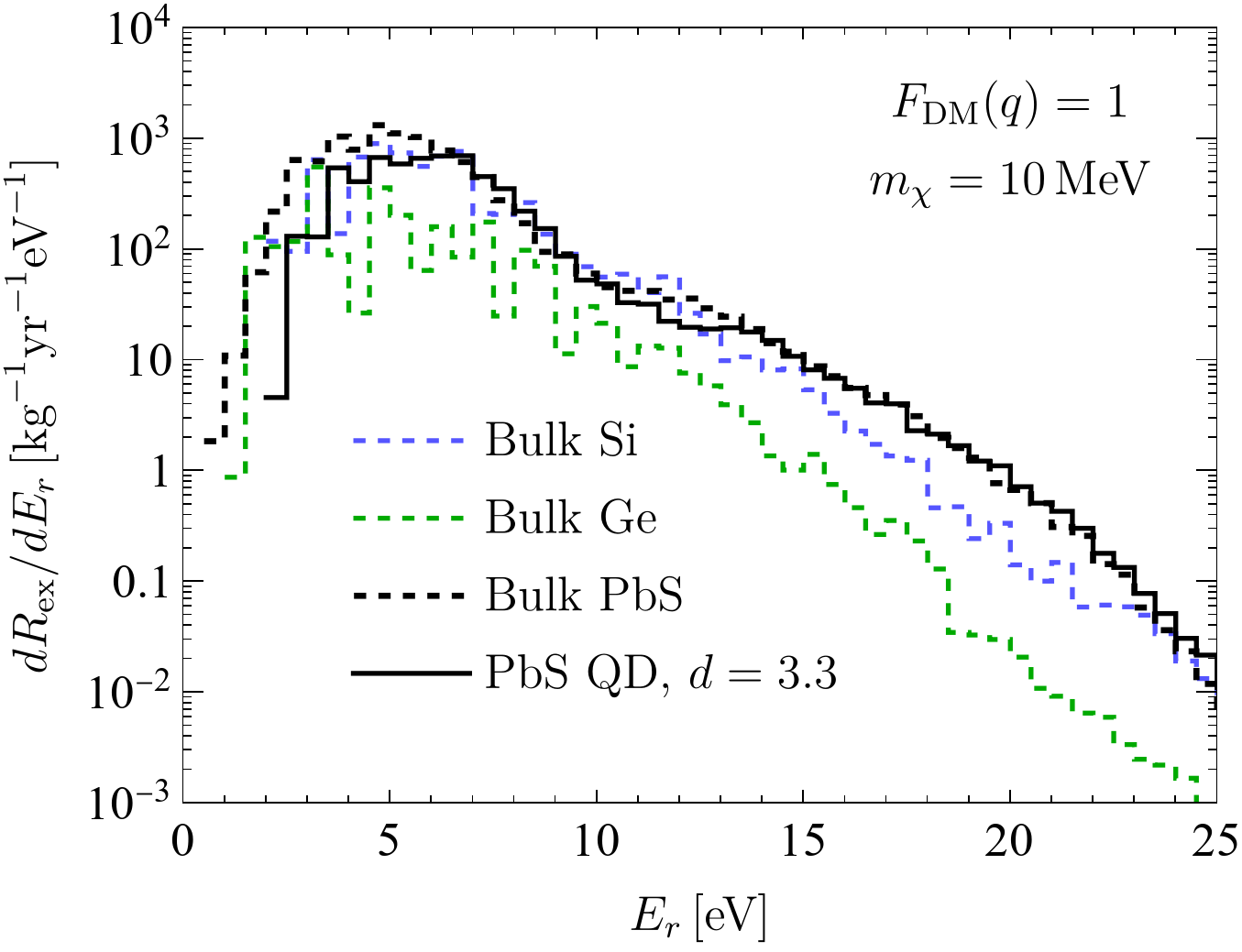}
    \includegraphics[width=0.45\textwidth]{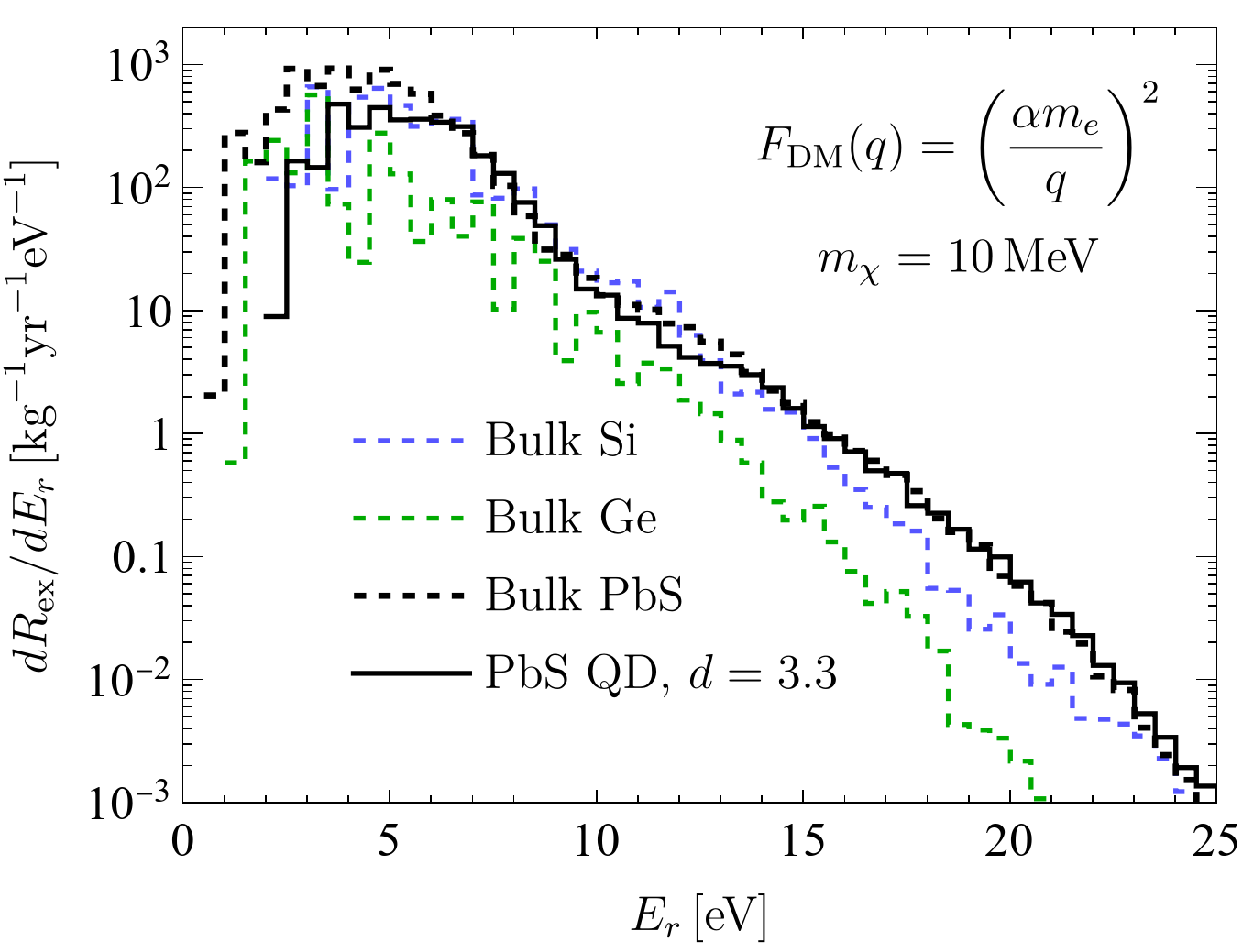}
    \caption{The differential excitation rate as a function of recoil energy for bulk Si (blue), bulk Ge (green), bulk PbS (dashed black), and PbS QDs with a diameter of d=3.3~nm (black). The normalized cross section is taken to be $\overline{\sigma}_e=10^{-37}~\text{cm}^2$ as a benchmark. Above the multi-exciton-generation (MEG) threshold ($E_{\rm th}\approx3.6$~eV), essentially all events are expected to induce at least two excitons. The left panel shows the rates assuming a momentum-independent DM form factor $F_{\rm DM}=1$, expected for models with heavy mediators, while the right panel shows the rates for a $F_{\rm DM}\sim1/q^2$, expected for models with light mediators.}
    \label{fig:10MeVrate}
\end{figure*}

Finally, $f_{\rm QD}(q,E_r)$ is the target form factor,
\begin{eqnarray}
|f_{\rm QD}|^2 & \equiv & \sum_{\rm fin}\sum_{\rm init} \delta(E_r - (E_{\kappa_f,n_f \ell_f m_f} - E_{\kappa_i,n_i \ell_i m_i})) \nonumber \\
& & \left \vert \int d^{3}r \,e^{-i\mathbf{q}\cdot\mathbf{r}} \Psi_{\kappa_f,n_f \ell_f m_f}^{*}(\mathbf{r})\Psi_{\kappa_i,n_i \ell_i m_i}(\mathbf{r}) \right \vert^2,
\label{eq:fqd}
\end{eqnarray}
where the sum is over populated initial states within a unit cell and all final states, and $E_{\kappa,n \ell m}$ are the energies of $\Psi_{\kappa,n \ell m}$. This form factor parameterizes the inelastic scattering probability of driving the transition $\{\kappa_i,n_i, \ell_i, m_i\} \rightarrow \{\kappa_f,n_f, \ell_f, m_f\}$, while imparting momentum $\mathbf{q}$.

As we discuss below, final states considered here are always sufficiently far from the band edge; the electronic structure of the QD is modeled using $\texttt{QE-dark}~$\cite{Essig:2015cda} as the equivalent bulk semiconductor, with a scissor correction applied to adjust the band gap to be $\Delta E_{\rm QD}(d)$. This correction essentially shifts the valence and conduction states of the bulk semiconductor such that the optical gap matches that predicted by $\Delta E_{\rm QD}(d)$. It is expected that for $q\gg2/d$ the interaction will be far more local than the size of the envelope of the edge states. Therefore, in addition to the arguments given above for the bulk-like electronic response above $E_r \gtrsim 2.5\;\text{eV}$, one expects that for $q \gg 0.5\; \text{keV}$ the $q$-dependence of the form factor will be dominated by the local bulk-like micro-physics encoded in $u_{\kappa}(r)$.

For calculations in this study we account for experimental irregularities in the QDs' size and surface which could alter the ideal $\Delta E_{\rm QD}$ in Eq.~\ref{eq:EgapQD}, by using the experimentally determined size-dependent expression for $\Delta E_{\rm QD}$~\cite{Greben2015},
\begin{align}
    \frac{\Delta E_{\rm QD}}{{\rm eV}} = \frac{\Delta E_g}{{\rm eV}} + \left( \frac{1}{0.0252 (d/{\rm nm})^{2}+0.283 (d/{\rm nm})^3}\right).
    \label{eq:QDexpEnergy}
\end{align} 
For the case of PbS, the bulk band gap is $\Delta E_g = 0.41\,\text{eV}$.

The numerical form factor is calculated using a plane-wave basis as in Ref.~\cite{Essig:2015cda}, 
 \begin{align}
    \vert f_{\rm QD}(q,E_{r})\vert^2 & = \frac{2\pi^2}{E_{r} \alpha m_e^2 V_{\rm cell}} \sum_{\kappa,\kappa^\prime}\int_{\rm BZ}\frac{V_{\rm cell}^2 d^3\mathbf{k}d^3\mathbf{k^\prime}}{(2\pi)^6} \\
    &\times E_r \delta(E_r - E_{\kappa^\prime}(\mathbf{k}^\prime)+E_{\kappa}(\mathbf{k}))
 \nonumber \\
    &\times \sum_{G^\prime} q \delta(q - |\mathbf{k}^\prime -\mathbf{k} +\mathbf{G}^\prime|) \vert f_{[\kappa,\mathbf{k},\kappa^\prime, \mathbf{k}^\prime,\mathbf{G}^\prime]} \vert^2\,,  \nonumber
\end{align}
with the function $f_{[\kappa,\mathbf{k},\kappa^\prime, \mathbf{k}^\prime,\mathbf{G}^\prime]}$ given by 
 \begin{align}
     f_{[\kappa,\mathbf{k},\kappa^\prime, \mathbf{k}^\prime,\mathbf{G}^\prime]}=\sum_{\mathbf{G}} \tilde{u}_{\kappa^\prime}^{*}(\mathbf{ G}^\prime+\mathbf{G}+\mathbf{k^\prime}) \tilde{u}_{\kappa}(\mathbf{G}+\mathbf{k})\,.
 \end{align}
In the above equations, $\mathbf{k}$ and $\mathbf{k}^\prime$ are the initial and final state momenta in the first Brillouin zone, $\mathbf{G}$, $\mathbf{G}^\prime$ are reciprocal lattice vectors, $V_{\rm cell}$ is the volume of the unit cell, and $\tilde{u}(\mathbf{k})$ is the $k$-space Fourier transform of $u(\mathbf{r})$,
 \begin{align}
     \tilde{u}_{\kappa}(\mathbf{k}) &= \frac{1}{\sqrt{V}}\int_{0}^{\infty} u_{\kappa}(\mathbf{r})e^{-i\mathbf{k}\cdot \mathbf{r}}  d^3\mathbf{r}\,.
 \end{align}
We take contributions from the 20 closest bands to the Fermi-level, with a kinetic energy cutoff of 272~eV (20~Ry), using norm-conserving non-relativistic pseudo-potentials for Pb and S.

Fig.~\ref{fig:10MeVrate} shows the calculated differential excitation rates for bulk Si ($\Delta E_{g}=1.1\; \text{eV}$), bulk Ge  ($\Delta E_{g}=0.67\; \text{eV}$), bulk PbS ($\Delta E_{g}=0.41\; \text{eV}$), and d=3.3~nm PbS QDs (for which $\Delta E_{\rm QD}=1.2\;\text{eV}$), for a DM particle mass of $m_\chi = 10$ MeV. As will be discussed below, the threshold energy for generation of MEG states is approximately $3 \times \Delta E_{\rm QD}$, i.e., approximately $3.6$ eV for the case of PbS. From the results of Fig.~\ref{fig:10MeVrate} it is evident that even at the relatively small DM mass of $10$~MeV, there is a significant rate well above this threshold energy. Therefore, for these examples the peak rate is expected to efficiently induce MEG states.

As will be discussed in Sec.~\ref{subsec:totalrate}, we consider signal events with electron recoil energies in the range $2.5 - 10$~eV. This choice is conservative, since it is sufficiently above the band edge so as not to be impacted by the strong confinement effects of the QD, and it is sufficiently small to avoid highly energetic final states, which will not emit fluorescence photons. Moreover, in this energy range, the calculation of the rate with $\texttt{QE-dark}~$\cite{Essig:2015cda} is expected to agree reasonably well with other approaches, which either include screening effects~\cite{Knapen:2021run,Hochberg:2021pkt,Knapen:2021bwg} or the effect of core electrons~\cite{Griffin:2021znd}. 

\subsection{Bosonic Absorption Rates}
\label{sec:QDabsorb}

In addition to scattering events, QDs are also sensitive to the absorption of bosonic DM by electrons (this was first considered for non-quantum-confined systems in~\cite{Pospelov:2007mp,An:2014twa}). For such interactions, the energy transfer to the target is equal to the entire DM particle mass. Therefore, since the typical threshold energies of QDs are in the $\sim $ eV range, this is also the natural DM mass scale to which such an experiment will be sensitive. Bosonic candidates such as scalars, pseudoscalars, and vectors (e.g., a dark photon) have previously been considered in the literature. However, we find sensitivity to unexplored parameter space---in particular regions that are not already ruled out by stellar cooling bounds---only for the case of dark photons. Therefore, we restrict our discussion to that model.

The dark photon, $A'$, is assumed to couple to the standard model photon, $A$, via the kinetic mixing portal, i.e.,
\begin{equation}
\mathcal{L} \supset \kappa F^{\mu \nu} F'_{\mu \nu} \,,
\end{equation}
where $ F^{\mu \nu}$ and $F'_{\mu \nu}$ correspond to the standard model and dark photon field strengths, respectively. Due to this simple structure, the dark photon absorption cross-section in any material can be computed from the empirical value of the standard model absorption cross-section through rescaling arguments, while taking into account in-medium effects. The rate for dark photon absorption per target atom is given by~\cite{Mitridate:2021ctr,Knapen:2021bwg}
\begin{equation}
    \sigma_{A'} v_{\rm DM}=\kappa^2 \frac{1}{|\epsilon|^2}\sigma_{A}\,.
\end{equation}
Here, $\sigma_{A}$ is the standard model photon absorption cross-section, $\epsilon$ is the dielectric function that captures the in-medium effects, and $v_{\rm DM}$ is the DM  velocity. Both $\sigma_{A}$ and $\epsilon$ can be obtained from experimental data; this is readily available for QDs with  $d=5$~nm (although not for $d=3.3$~nm). We take  $\sigma_{A}$ from Ref.~\cite{jahromi2019lead} for $m_{A'}\lesssim 5.5~\textrm{eV}$ and from Ref.~\cite{kanazawa1998optical} for higher masses. While these were measured for bulk PbS, this is justified for QDs, since QDs are expected to have bulk-like behavior for absorption above $\sim 2.5~\textrm{eV}$~\cite{moreels2009size}. The dielectric function is obtained from Ref.~\cite{hechster2017modeling}.

\subsection{Excitations of Quantum Dots}
\label{subsec:Excitations}

If the QD is initially in its ground state and an amount $\Delta E_{\rm QD}$ or more of energy is transferred in an interaction event, an electron-hole pair is created. If the energy transfer is below $2 \times \Delta E_{\rm QD}$, a single excitonic final state can be created, which has the potential to decay into a detectable photon. Theoretically, if the transferred energy exceeds $2 \times \Delta E_{\rm QD}$, a multi-exciton state is possible. However, it has been shown experimentally that for Pbs (and for PbSe), the threshold for MEG is closer to $3 \times \Delta E_{\rm QD}$~\cite{midgett2013size,schaller2004high,ellingson2005highly,hardman2011electronic}.

It has also been shown that the surface treatment of the QD, as well as the solvent used to suspend the QDs, play important roles in optimizing the quantum yield of both single exciton and multi-exciton states~\cite{beard2009variations}. The number of band-edge excitons generated per primary excitation, i.e., the quantum yield ${\rm QY}_{\rm ex}$, is energy-dependent, and experimentally the quantum yield for PbS (and PbSe) is typically optimized to be around ${\rm QY}_{\rm ex} \approx 2-3$ at $E_r = 4 \times \Delta E_{\rm QD}$~\cite{midgett2013size,schaller2004high,ellingson2005highly,hardman2011electronic,beard2009variations}. Since the rate of non-radiative Auger recombination, whereby an electron-hole pair recombines and imparts its energy into a third carrier, scales as $n_{e/h}^3$ ($n_{e/h}$ is the carrier density), final states with more than two excitons do not live long enough to emit multiple photons.

\begin{figure}
    \includegraphics[width=0.45\textwidth]{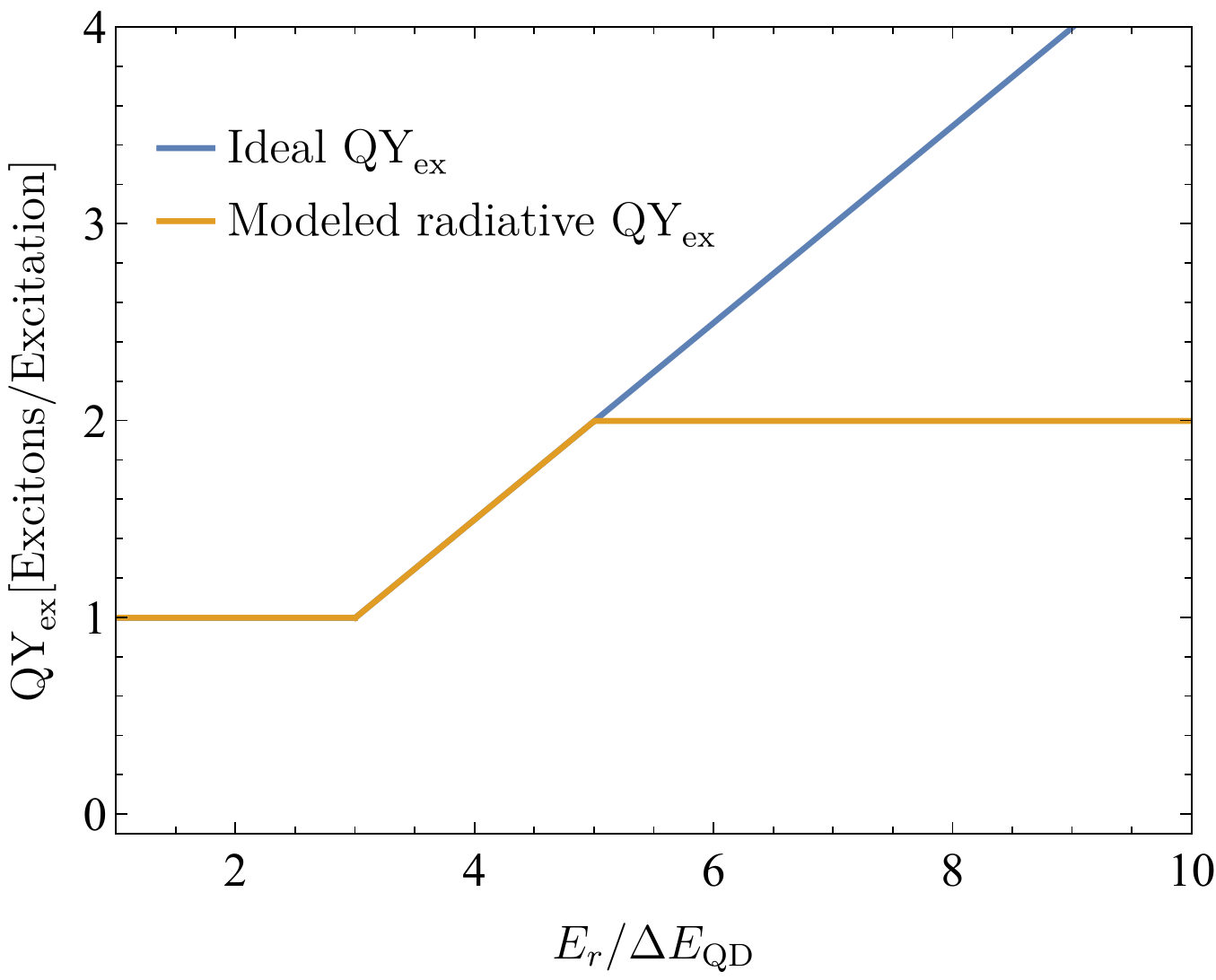}
    \caption{The exciton response of PbS. The blue line represents the theoretical exciton quantum yield of PbS QDs~\cite{midgett2013size,schaller2004high,ellingson2005highly,hardman2011electronic}. In order to account for Auger recombination, the modeled rediative response is taken to saturate when two excitons are created (orange line), since Auger recombination will quickly deplete higher exciton multiplicities.}
    \label{fig:QY_ex}
\end{figure}

To remain conservative, we assume that all multi-exciton final states with a number of excitons greater than three immediately decay to a bi-exciton state before emitting photons. We model the function QY$_{\rm ex}$ as shown in orange in Fig.~\ref{fig:QY_ex}. Below $3 \times \Delta E_{\rm QD}$, the exciton quantum yield is unity, corresponding to single exciton generation, while above this value the function is modeled as a linear response up to 2 excitons at $5 \times \Delta E_{\rm QD}$ and beyond which no additional excitons are created. This functional form roughly captures the spread of values found in the literature with a slight bias towards the conservative direction. Note that QY$_{\rm ex}$ should be thought of as an experimental measure of the mean number of excitons created in each excitation and it can therefore be a continuous response. Also shown in blue in the figure is the ideal theoretical MEG quantum yield for PbS, which grows linearly with $E_r$.

\subsection{Photon Generation and Collection}
\label{subsec:Photon_Collection}

Once single or multiple band-edge excitons are generated, they decay either radiatively by fluorescing, or non-radiatively via, e.g.,~collisional quenching, Auger recombination, or phonon emission. As discussed above, single-exciton fluorescence is highly efficient with a quantum yield of order QY$_{\gamma} \approx 0.5$~\cite{yang2017iodide} while multi-exciton fluorescence competes with Auger recombination, which lowers the quantum yield for such multi-photon emission. For example, the bi-exciton, two-photon emission has a quantum yield of order  QY$_{\gamma\gamma} \approx 0.05$~\cite{beyler2014sample}.

Since we make the conservative assumption that any exciton multiplicity greater than two is rapidly reduced to the bi-exciton state with no intermediate photon emission, the resulting signal consists of either a single photon or two coincident photons. For the latter case, the coincidence time is of order the bi-exciton lifetime, $\tau_{\gamma\gamma} = \tau_{\gamma} \times \text{QY}_{\gamma\gamma} / 4 \approx 12 - 65$ ns, where $\tau_{\gamma} \approx 1-5 \, \mu$s is the single-exciton radiative lifetime for PbS QDs (the same is true for PbSe)~\cite{mangum2013disentangling,moreels2009size}.

The one and two-photon differential signal rates are given by 
\beq
\frac{dR_{\rm ph}}{d\ln{E_r}} = \frac{dR_{\rm ex}}{d\ln{E_r}}
    \begin{cases} 
      \text{QY}_{\rm ex} \cdot \xi \cdot \text{QY}_{\rm BF} & ,1\gamma \\
      (\text{QY}_{\rm ex}-1) \cdot \frac{\text{QY}_{\gamma\gamma}}{2} \cdot \left(\frac{\xi \cdot \text{QY}_{\rm BF}}{\text{QY}_{\gamma}}\right)^2 & ,2\gamma
   \end{cases}
    \label{eq:sigRate}
\eeq
where $\xi$ is the photon detection efficiency of the photodetector, which we take to be $\xi=0.25$. This is a typical value for detection efficiency in PMTs used in scintillating DM direct detection experiments~\cite{Blanco:2019lrf,blanco2020dark}, although newer technologies such as Superconducting Nanowire Single Photon Detectors (SNSPDs), Microwave Kinetic Inductance Detectors (MKIDs), and Transition Edge Sensors (TESs) aim to push this detection efficiency to near unity. The bulk fluorescence quantum yield, QY$_{\rm BF}$, is the probability of a single band-edge exciton to emit a photon that can travel macroscopic distances without being reabsorbed by other chromophores (optically active targets). This is an experimental measure of optical losses in the suspension volume (we assume that other losses, such as from photons being absorbed by the side walls, are negligible; see also discussion in ref.~\cite{Collar:2018ydf}). The value of QY$_{\rm BF}$ is discussed in detail below. QY$_{\rm BF}$/QY$_\gamma$ is the self-absorption normalized by the probability of emitting a single photon from a single band-edge exciton. The factor in the second parentheses of the two-photon signal is the detectability of a single-photon event. Note that the factor of $1/2$ in the two-photon signal accounts for requiring that each of the two photons hits opposing photodetectors.

As discussed in App.~\ref{sec:QY}, the value of QY$_{\rm BF}$ (and therefore the reabsorption probability) is correlated with the Stokes shift of the emission and absorption spectra, with the width of the emission spectrum, and with the concentration of QDs. The width of the emission spectrum is dominated by the dispersion of the size distribution of QDs, where a tighter size distribution improves fluorescence quantum yield. At room temperature, the line width of a single QD is on the order of 100~meV while that of typical colloidal ensembles is about 250~meV~\cite{peterson2006fluorescence}. Experimentally, it has been shown that reabsorption of photons becomes significant for such colloidal suspensions above QD particle concentrations of $C_{\rm part,QD} \approx 10\;\mu\text{M}/\text{L}$~\cite{Greben2015}. Therefore, in all calculations presented in this study, we take this maximal concentration value, from which one can easily calculate the density of target material given in Eq.~\eqref{eq:N_t}. 
In particular,
\begin{align}
    \rho_{\rm QD} & = C_{\rm part,QD} \cdot \rho_{\rm semi} \cdot \frac{4\pi}{3} \left( \frac{d}{2}\right)^3 \label{eq:C_m} \\
    &\approx 2.5 \; \text{g}/\text{L} \left( \frac{C_{\rm part,QD}}{10\;\mu\text{M}/ \text{L}} \right) \left( \frac{\rho_{\rm semi}}{6.29 \text{g}/\text{cm}^{3}} \right)\left( \frac{d}{5 \text{nm}} \right)^3 \nonumber,
\end{align}
where $\rho_{\rm semi}$ is the bulk semiconductor density. It should be noted that $\rho_{\rm QD}$ can potentially be increased up to $22$~g/L for $7.5$~nm PbSe QDs at a concentration of $C_{\rm part,QD} \approx 20\;\mu\text{M}/ \text{L}$. From Eq.~\eqref{eq:C_m} it is evident that even with a very simple setup of a $\mathcal{O}$(L) instrumented target volume, such as that of Ref.~\cite{Blanco:2019lrf}, effective masses of $\mathcal{O}(10\text{ grams})$ of semiconducting scintillator are achievable.

Under the assumption of this close-to-maximal concentration, we adopt a conservative room temperature value of QY$_{\rm BF} = 0.2$ also based on experimentally determined values (which can in fact be as large as QY$_{\rm BF} = 0.25$ for PbS QDs with $d=3-3.3$ nm~\cite{Greben2015}). However, we note that the photoluminescent quantum yield of PbS QDs improves significantly at lower temperatures, reaching values more than twice that of room temperature QDs already at 100~K~\cite{caram2016pbs}.  Additional details regarding QY$_{\rm BF}$ are presented in App.~\ref{sec:QY}.

\subsection{Total Event Rate}\label{subsec:totalrate}

In order to calculate the sensitivity of a hypothetical DM experiment, one quantity of interest is the total, directionally averaged rate. Since a continuously resolved energy spectrum is not measurable for the envisioned setup, the differential rate must be integrated over the energy-domain included in the experimental signal. The total single- or two-photon rate is given by 
\beq
    R_{\rm ph} = \int_{E_{\rm th}}^{E_{\rm cutoff}} \frac{dR_{\rm ph}}{d\ln{E_r}} \frac{dE_r}{E_r}\,.
\eeq
The lower limit of integration depends on the threshold of the target for the specific signal of interest. We conservatively take $E_{\rm th}$ to be the threshold for MEG, namely $3\Delta E_{\rm QD}\simeq 3.6$~eV, for both the single- and two-photon signals. For the case of MEG, this threshold is experimentally measured and is well within the region where the PbS QDs behave like bulk PbS. For the case of the single-photon signal, we remain conservative by not counting the region near the band edge where strong confinement effects might affect the optical and electronic excitation response of the QDs. Including this near-bandgap region below the MEG threshold would serve to increase the calculated rate and improve the projected sensitivity of the setup. However, from Fig.~\ref{fig:10MeVrate} we infer that such an improvement would not be significant unless the DM mass is close to threshold. Note that the noisy behavior apparent in the low energy region of the $dR_{\rm ex}/d E_r$ curves shown in Fig.~\ref{fig:10MeVrate} (which arises from numerical uncertainties in $\texttt{QE-dark}$ close to the band edge) are mostly removed because of this choice for $E_{\rm th}$. The upper limit is set by the highest energy expected to produce a visible signal; for higher energies, processes such as ligand excitation or QD ionization could prevent excited states from relaxing to photo-luminescent final states. For our calculations, we take the maximum recoil energy to be $E_{\rm cutoff} = 10$~eV as a conservative choice and note that our results do not depend strongly on this choice.

\section{Experimental Setup and Background Rates}
\label{sec:BGs}

For results presented in this study, we consider two classes of experimental setups: 1) a setup based on currently available and mature photodetectors such as PMTs, and 2) a more futuristic setup based on high sensitivity photodetectors such as SNSPDs. The former option is achievable on relatively short timescales and therefore, for this setup, we present results only for a coincident two-photon signal under the assumption that this low-background option is a likely first iteration of the experimental concept. The latter option is likely achievable only on longer timescales. For this futuristic setup, we present results for the more challenging single-photon signal, and we show results for an idealized zero background setup.  While achieving zero background rates is likely to be experimentally very challenging, we present such results in order to show the ultimate sensitivity of the setup.

For the case of PMTs, previous studies focusing on liquid scintillators have proposed and deployed large-area single PMTs as photodetectors~\cite{Blanco:2019lrf,blanco2020dark,Collar:2018ydf} to instrument liter-scale scintillation volumes. These studies suggest that background rates of about $1$~Hz for $5$ inch diameter bialkali PMTs should be easily obtained when running with moderate overburden ($6.25$ meters water equivalent), and $0.1$~Hz should be possible in deep underground locations and with ideal thermal control. Note, that in previous studies this background was measured to be the dark-count reduced steady-state background in a differential measurement, i.e., data taken with the PMT blind vs observing the scintillator volume. Similar differential measurements can be done to reduce backgrounds here. Furthermore, we note that GaAs photocathodes have been measured to have 1~Hz dark counts at temperatures of $-40^\circ$~C, while side-on bialkali photocathodes reach approximately 0.3~Hz at around $-20^\circ$~C~\cite{hamamatsu2007photomultiplier}. We abstain from making specific recommendations about the exact geometry and composition of the PMT but note that technology exists that can meet the necessary benchmarks mentioned in this study. Furthermore, we point out that manufacturing variance allows hand-picked units to outperform the mean manufacturer specification for dark-counts (see, e.g., Ref.~\cite{Collar:2018ydf}).

In the calculations that follow, we assume PMT background rates and show results for both $\sim$1~Hz and $0.1$~Hz. For the case of multiple coincident photons we discuss the dominant sources of backgrounds which are expected to be blackbody radiation and the dark counts of the PMT detectors themselves. However, additional backgrounds, such as Cherenkov radiation, transition radiation, and low-energy luminescence~\cite{Du:2020ldo} from the solvent and from any dielectric holders and other structures inside the detector vessel could also contribute. These effects need to be evaluated carefully before confronting a positive signal with a DM interpretation. For the case of more sensitive photodetectors such as SNSPDs, the detector dark counts will be lower and environmental backgrounds will likely dominate. Furthermore, we note that PMTs in the arrangement presented in Fig.~\ref{fig:QDdiagram} could detect each other's dynode glow during a photoavalanche. This background is straight-forward to reduce. For example, since the dynode emission from a PMT has a broad bandwidth, while the fluorescence photons from the QDs are essentially monochromatic, a simple notch filter can be used to ameliorate any potential dynode-glow background. This technique may also be used to reduce potential direct Cherenkov backgrounds. In either case, in-situ characterization would be necessary. 

The single photon background rate of a single photodetector with surface area, $A_{\rm det}$, from blackbody radiation can be determined via~\cite{Essig:2019kfe},
\begin{equation}
    R_{\rm BBR} = \frac{\Delta \omega \, \omega^2}{\pi^2} A_{\rm det} e^{-\frac{\omega}{T}}\,.
\end{equation}
Here, $\omega$ is the energy of the photon, $\Delta \omega$ is the detector energy resolution, and $T$ is the ambient temperature. For $\Delta \omega \approx \omega =1.2~\textrm{eV}$ and $A_{\rm det} = \pi \, (2.5~\textrm{inch})^2$, one finds that $T \lesssim 230$ K reduces $R_{\rm BBR}$ to be below $1$~Hz.

The envisioned setup includes modules of target material, each as shown in Fig.~\ref{fig:QDdiagram} and each including a single cylindrical tank instrumented with two photodetectors. A two coincident photon signal occurs when both photodetectors record hits within some coincidence time, $\Delta t_{\rm co}$. The background rate for such coincident events can be calculated from the Poissonian single-photon background rate of each detector, $R^{\rm BG}_{\gamma}$ (assumed to be the same for both detectors). As long as $R^{\rm BG}_{\gamma} \Delta t_{\rm co} \ll 1$, the result for the two-coincident-photon rate is 
\beq
R^{\rm BG}_{\gamma\gamma} \approx 2 (R^{\rm BG}_{\gamma})^2 \Delta t_{\rm co}\,.
\label{eq:f2ph}
\eeq
Taking $R^{\rm BG}_{\gamma} = 1$~Hz and $\Delta t_{\rm co} = 100$~ns (a few times the lifetime of the two-exciton state), one finds that for an exposure time of $t_{\rm exp} \approx 1$~month, the number of coincident two-photon events is $R^{\rm BG}_{\gamma\gamma} t_{\rm exp} \approx 0.52$. Since the instrumentation described here corresponds to approximately $10$ grams of target material (see Sec.~\ref{subsec:Photon_Collection}), a setup with an exposure of order $10$~g-months is expected to be essentially background free for the coincident two-photon signal and any additional reduction of $R^{\rm BG}_{\gamma}$ would do little to the overall experimental sensitivity.

In order to achieve larger exposures, we propose simply running in parallel several replica setups shown in Fig.~\ref{fig:QDdiagram}. Increasing exposure in this way features a linear scaling of the background rate with the target volume, as opposed to the less advantageous scaling of background rates expected from instrumenting a larger tank. Thus, for example, one could envision running ten detectors for one year and achieving exposures of order $100$~g-years with $R^{\rm BG}_{\gamma\gamma}$ only ten times larger than the $10$~g-month case. Results below are presented for these two example exposures, $10$ g-months and $100$ g-years, and for single photon background rates of $R^{\rm BG}_{\gamma}=1$ and $0.1$~Hz, for the coincident two-photon signal. For the futuristic single-photon signal, results are presented for an exposure of $100$ g-years and a zero background rate.

The ($1-\alpha$)\% C.L.~signal for an exposure time, $t_{\rm exp}$, for a two-photon (single-photon) background rate, $R^{\rm BG}_{(\gamma)\gamma}$, is given by,
\beq
     R_{(1-\alpha)\%} = \frac{1}{2} \frac{\chi^2}{t_{\rm exp}} \left(2(R^{\rm BG}_{(\gamma)\gamma} t_{\rm exp} +1),1-\frac{\alpha}{2}\right)\,, 
     \label{eq:poissonCL}
\eeq
where $\chi^2$ is the $(1-\frac{\alpha}{2})$-th quantile of the chi-squared distribution of $2(R^{\rm BG}_{(\gamma)\gamma} t_{\rm exp} +1)$ degrees of freedom. For example, for a background-free exposure, i.e., $R^{\rm BG}_{(\gamma)\gamma} t_{\rm exp} \ll 1$, the 90\% C.L. number of events is $R_{90\%} t_{\rm exp} = 2.99$. For a $10$~g-month exposure and a single-photon background rate of $R^{\rm BG}_{\gamma} = 1$~Hz ($0.1$~Hz), this corresponds to a total coincident two-photon background rate of 0.52 ($5.2\times 10^{-3}$) events / month. This is essentially background free and corresponds to 90\% C.L. total event counts of 3.95 (3.01). For a $100$~g-year exposure, the total coincident two-photon background rate is 63.1 ($0.63$) events / year. This corresponds to 90\% C.L. total event counts of 77.8 (4.13).

The calculations above have focused on the case of PMTs as photodetectors. Ideally however, low-background photodetectors such as SNSPDs, MKIDs, or TESs could be used for photon counting. However, scaling up the effective collection area of these devices remains challenging. Skipper-CCDs present another interesting photodetector candidate in the context of scintillator detectors, and one could envision having two CCDs read photons coming from a target; the CCDs have intrinsically low backgrounds, although further reduction of the time to read the entire CCD may be needed for a two-photon coincidence signal counting.\footnote{Taking the Skipper-CCDs used by SENSEI~\cite{SENSEI:2020dpa} as an example, one option is to collect the charge of the 5.4~million pixels of the CCD into a single pixel and sample that pixel 300 times (to achieve single-electron resolution); this can be done in about 0.26~s.  Taking a single-electron background rate of $10^{-4}$~electrons/pixel/day, achieved in~\cite{SENSEI:2020dpa} in a detector operating inside a small shield about 100~m underground, one finds a two-photon-coincident background signal rate for two Skipper-CCDs operating in parallel of $\sim$640 events/year. This is higher than the two-photon-coincident background rate from two PMTs operating with a 1~Hz background rate; however, the readout time scales linearly with the number of amplifiers and the photon detection efficiency is $\mathcal{O}$(1) compared to $\xi=0.25$ assumed for the PMTs. Hence Skipper-CCDs should be considered as a possible photodetector choice.}

\section{Dark Matter Sensitivity}
\label{sec:sens}
 
\subsection{DM-Electron Scattering}
\label{subsec:Scattering}

 \begin{figure*}
    \centering
    \includegraphics[width=0.45\textwidth]{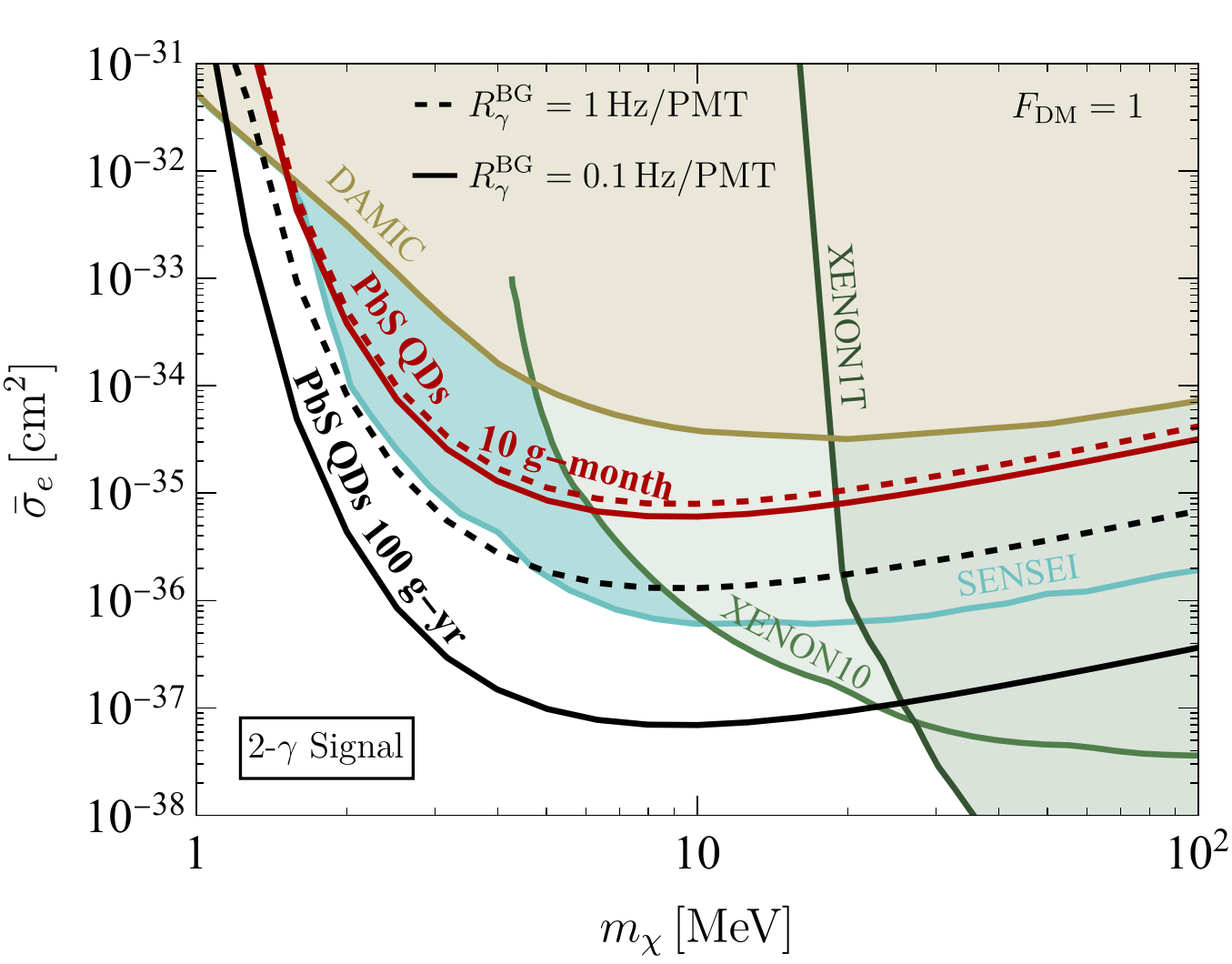}
    \includegraphics[width=0.45\textwidth]{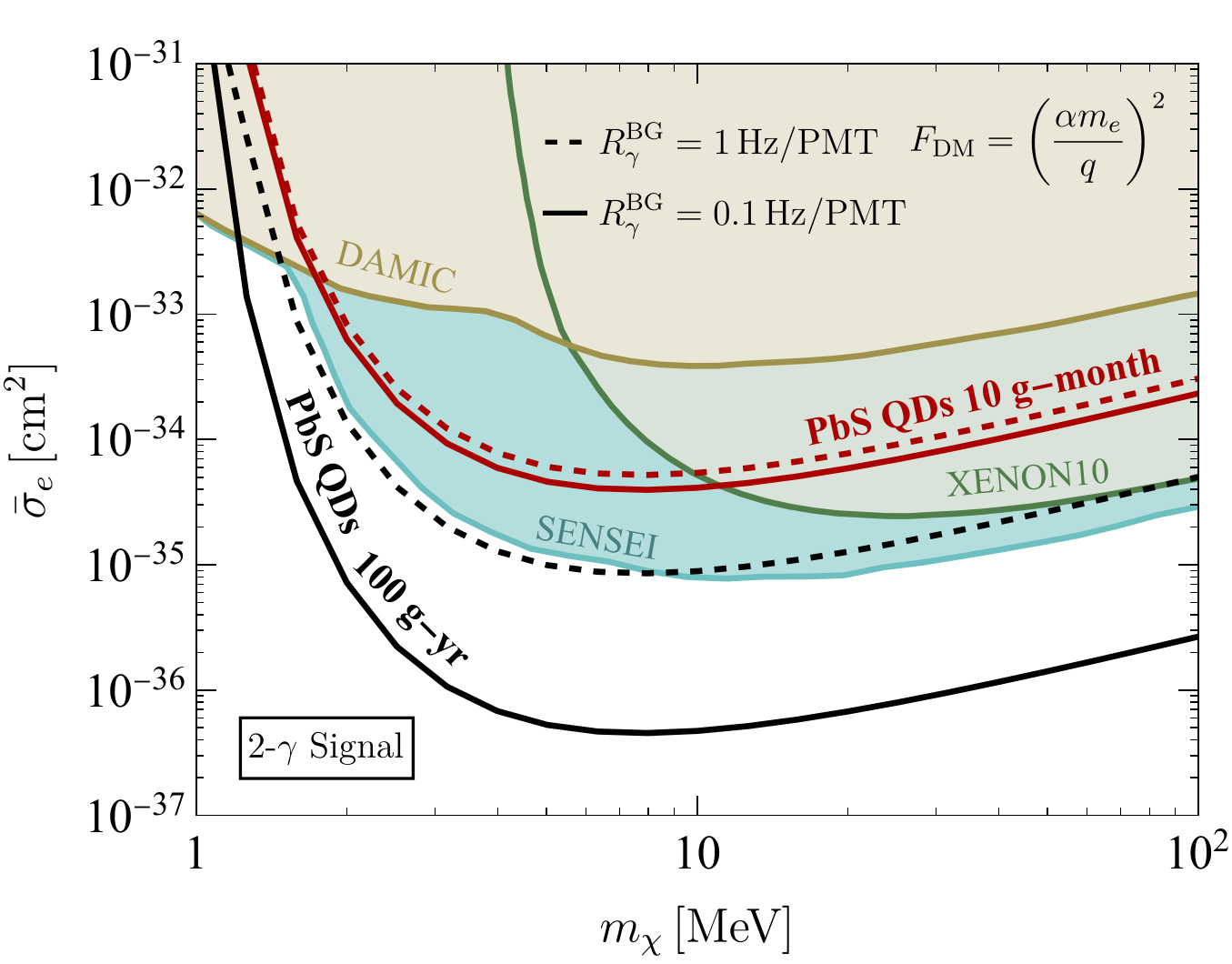}\\
    \includegraphics[width=0.45\textwidth]{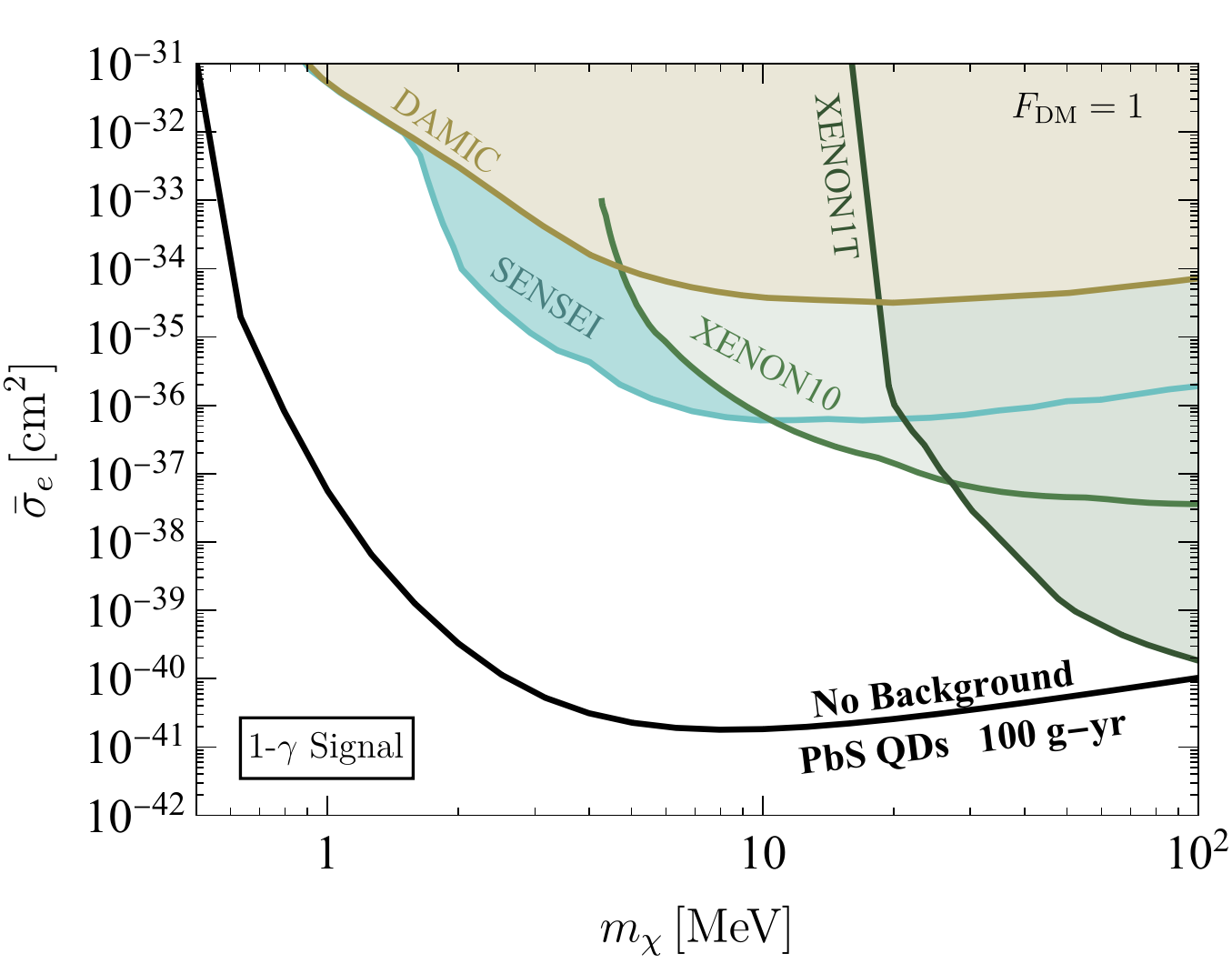}
    \includegraphics[width=0.45\textwidth]{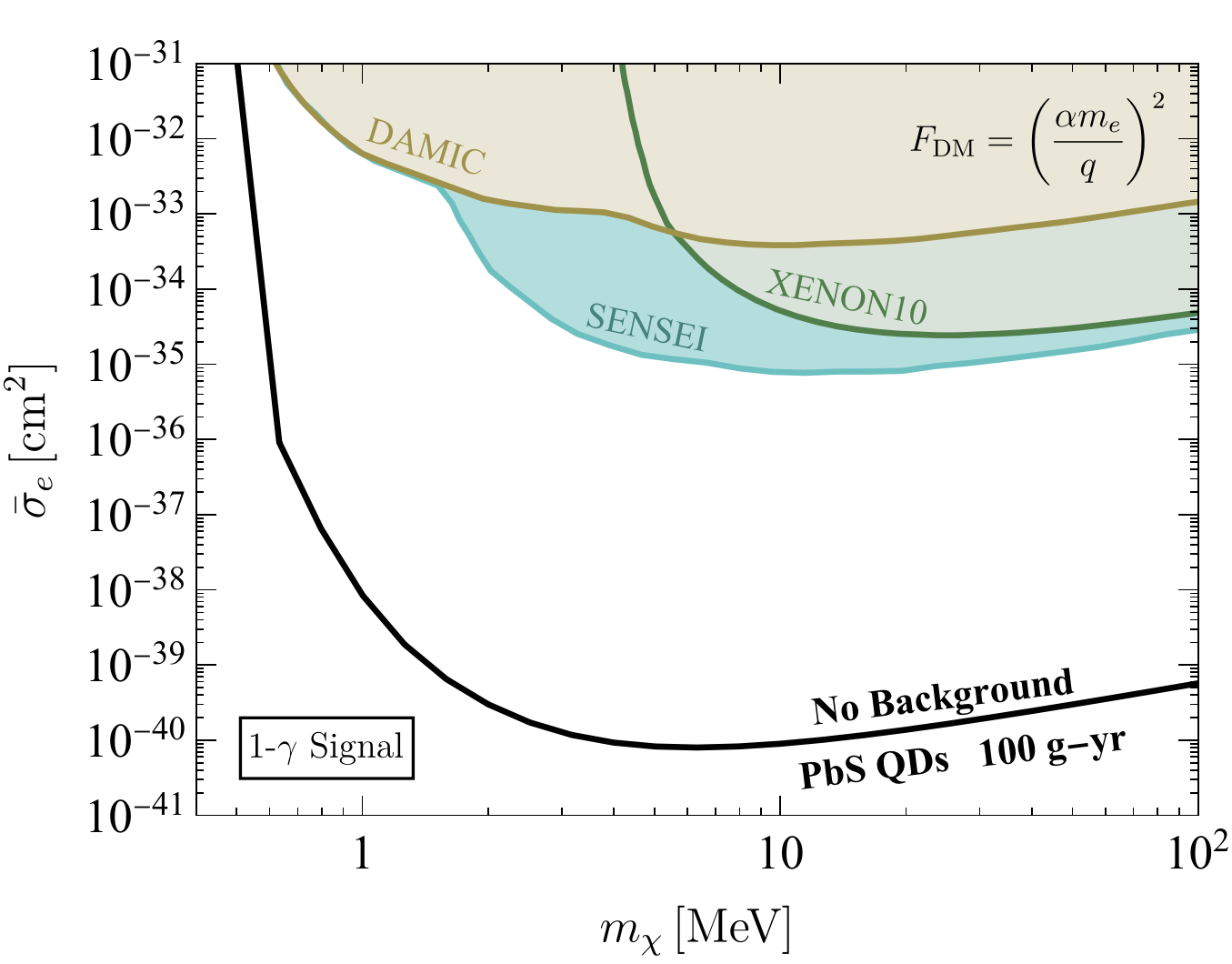}
    \caption{The 90\% C.L.~sensitivity for scattering events with a $10\,$g-month ($100\,$g-year) exposure corresponding to a $1$~L ($10\times 1$~L) colloidal suspension of $d=3.3\,$nm PbS QDs is shown in red (black). The dashed lines are for a steady-state background rate of $1\,\text{Hz}/\text{PMT}$, while the solid lines are for a steady-state background rate of $0.1\,\text{Hz}/\text{PMT}$,  assuming this background scales linearly with target volume. Shaded regions represent existing constraints from SENSEI~\cite{Barak:2020fql}, XENON10~\cite{Essig:2017kqs}, XENON1T~\cite{Aprile:2019xxb}, and DAMIC~\cite{DAMIC:2019dcn}. The top frames show the the calculated sensitivity when searching for a two-photon coincident signal over a steady-state single-photon background of $R_{\gamma}^{\text{BG}}=1$~Hz ($R_{\gamma}^{\text{BG}}=0.1$~Hz) as shown by the dashed (solid) lines. The bottom panels show the calculated sensitivity of a futuristic 100 g-yr QD exposure when searching for a single-photon signal assuming no backgrounds. The left (right) panels show these sensitivities and exclusions as calculated assuming DM form factors of $F_{\text{DM}}=1$ ($F_{\text{DM}}\sim1/q^2$), appropriate for models with heavy (light) mediators.}
    \label{fig:QDsense}
\end{figure*}

Fig.~\ref{fig:QDsense} shows the expected sensitivity of the proposed setup to the reference cross section, $\bar{\sigma}_e$, for DM-electron scattering events, for the coincident two-photon signal (top panels) and for the single photon signal (bottom panels) for a maximally concentrated solution of $d=3.3\,$nm PbS QDs. For the coincident two-photon signal, results are shown for $10\,$g-month and $100\,$g-year exposures and backgrounds are calculated assuming $R_\gamma^{\rm BG} = 1$~Hz (dashed curves) and $R_\gamma^{\rm BG} = 0.1$~Hz (solid curves). This setup should be achievable on relatively short timescales. For the single-photon signal, results are shown for a more futuristic $100\,$g-year exposure under the assumption that single-photon backgrounds are reduced to be negligible. For both signal types, results are shown for a contact interaction corresponding to $F_{\rm DM} = 1$ (left panels) and for the case of an ultralight (or massless) mediator corresponding to $F_{\rm DM} = (\alpha m_e)/q^2$ (right panels). All results are compared to existing constraints on DM-electron scattering from SENSEI~\cite{Barak:2020fql}, XENON10~\cite{Angle:2011th,Essig:2012yx,Essig:2017kqs}, XENON1T~\cite{Aprile:2019xxb}, and DAMIC~\cite{DAMIC:2019dcn} (see also~\cite{Blanco:2019lrf,EDELWEISS:2020fxc,SuperCDMS:2018mne}). 

Since the expected number of background events is negligible for a $10$~g-month exposure, the sensitivity is driven solely by the signal generation efficiency of the target mass. Because this pilot experiment is background free, it is ideally suited for the characterization of the detector and naturally lends itself as a scale that could be deployed as an experimental proof-of-concept. Even at this very conservative size, the sensitivity of such an exposure is competitive with existing constraints by SENSEI~\cite{SENSEI:2020dpa} at DM masses below 10~MeV where it would already improve on existing bounds by DAMIC at SNOLAB~\cite{DAMIC:2019dcn}. Lowering the operating temperature in order to improve the photoluminescence quantum yield and reduce PMT noise, optimizing the QD surface chemistry to improve the bi-exciton radiative quantum yield, or reducing the rate of Auger recombination at this scale could already probe new parameter space.

We find that for the coincident two-photon signal, exposures of order $100$~g-years and single-photon background fluxes of order $0.1$~Hz probe parameter space significantly beyond the existing SENSEI constraints~\cite{SENSEI:2020dpa}. The more futuristic setup, which we envision as being sensitive to single-photon events with negligible backgrounds, is expected to probe many orders of magnitude of new parameter space for both DM form factors.

\begin{figure*}[htbp]
    \centering
    \includegraphics[width=0.45\textwidth]{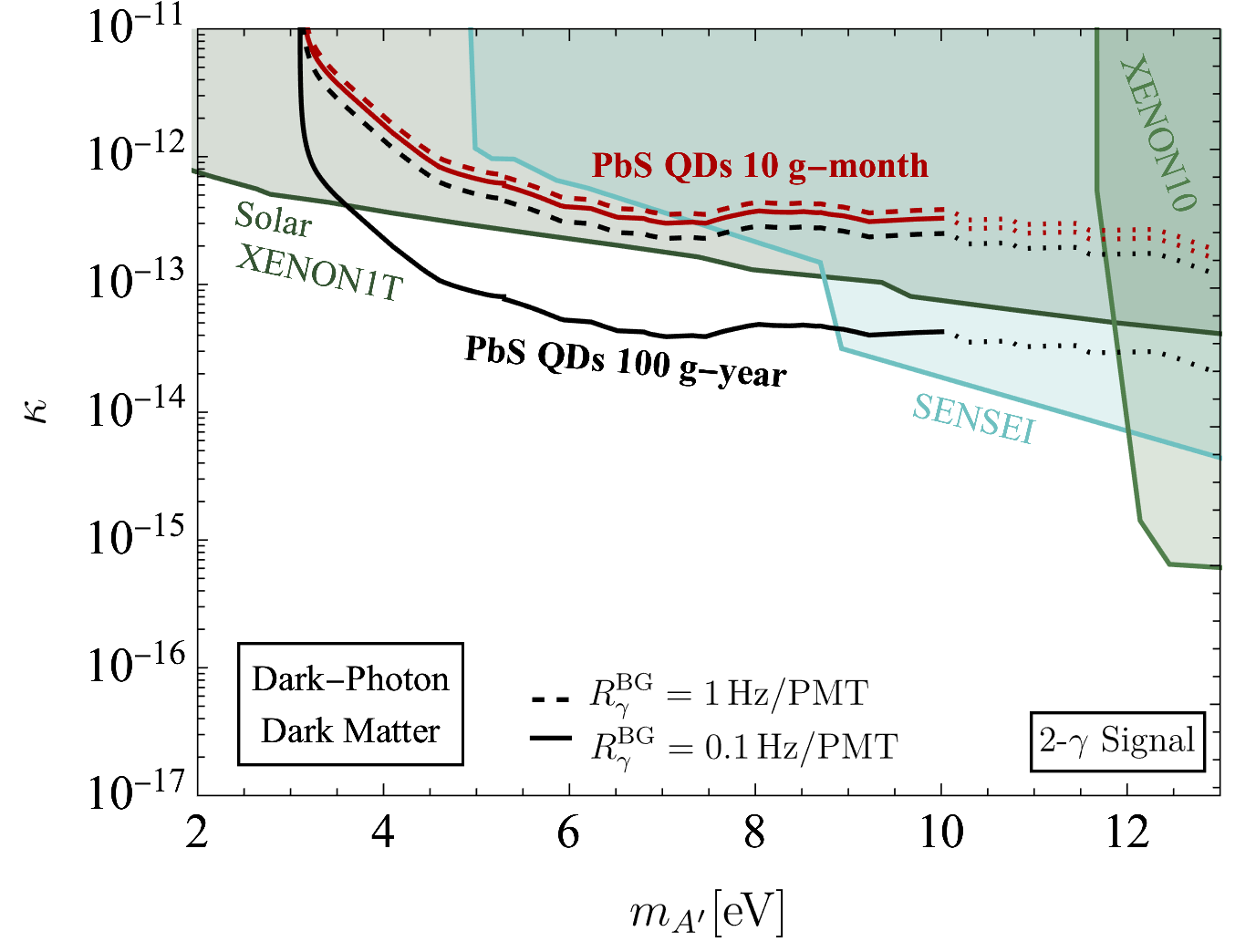}
    \includegraphics[width=0.45\textwidth]{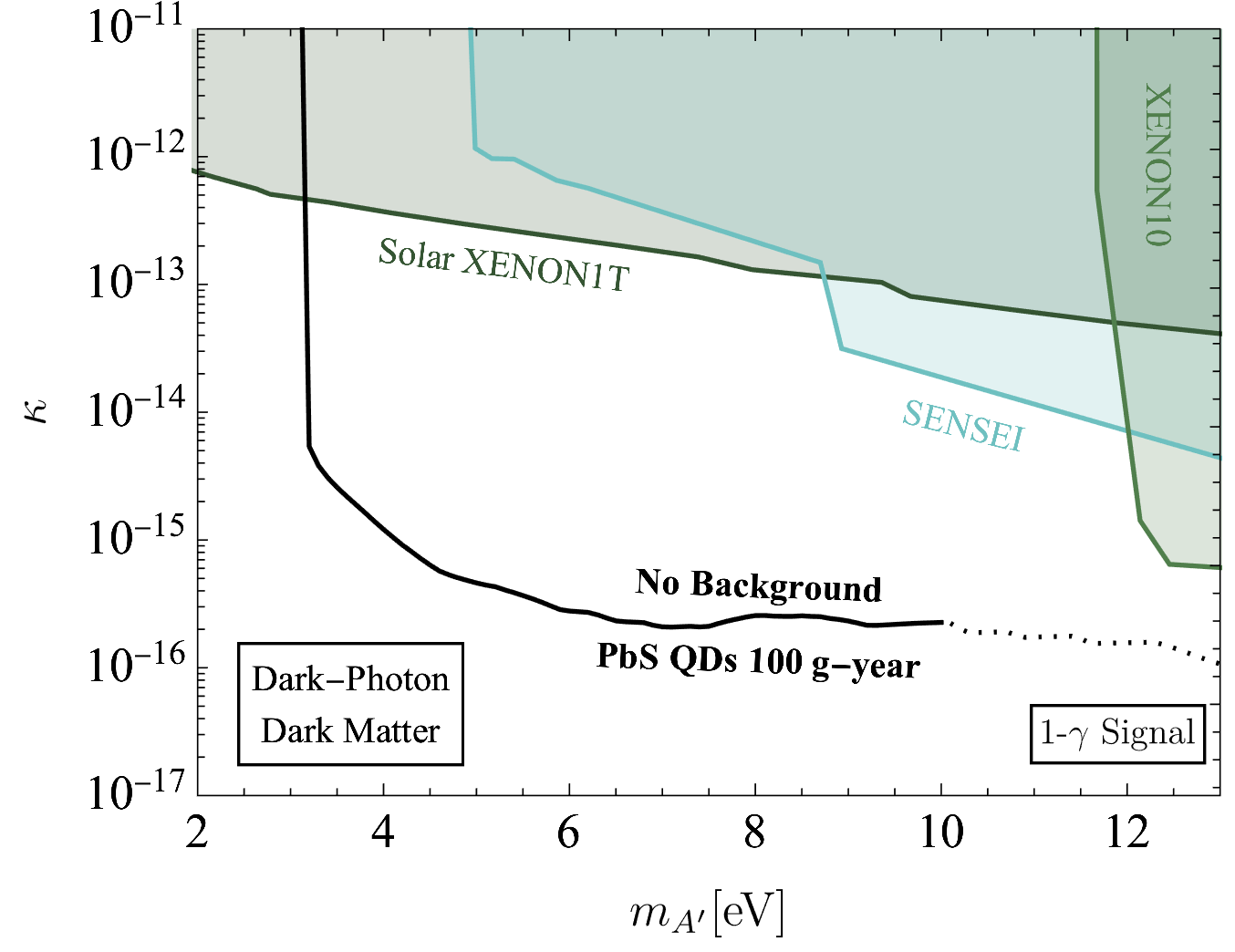}
    \caption{The 90\% C.L.~sensitivity for absorption events for dark photon dark matter with a $10\,$g-month ($100\,$g-year) exposure corresponding to a $1$~L ($10\times 1$~L) colloidal suspension of $d=5\,$nm PbS QDs is shown in red (black). The dashed lines are for a steady-state background rate of $1\,\text{Hz}/\text{PMT}$, while the solid lines are for a steady-state background rate of $0.1\,\text{Hz}/\text{PMT}$,  assuming this background scales linearly with target volume. Shaded regions represent existing constraints from SENSEI~\cite{SENSEI:2020dpa} modified to include in-medium effects (see text for details), XENON10~\cite{Bloch:2016sjj}, and  XENON1T~\cite{An:2020bxd}. The left panel shows the the calculated sensitivity when searching for a two-photon coincident signal over a steady-state single-photon background of $R_{\gamma}^{\text{BG}}=1$~Hz ($R_{\gamma}^{\text{BG}}=0.1$~Hz) as shown by the dashed (solid) lines. The right panel shows the calculated sensitivity of a futuristic 100 g-yr QD exposure when searching for a single-photon signal assuming no backgrounds. All lines are dotted above $m_{A'}=10~\textrm{eV}$ to indicate that for these masses the response of the QD is uncertain.}
    \label{fig:abs}
\end{figure*}

\subsection{DM Absorption by Electrons}
\label{subsec:Absorption}

Fig.~\ref{fig:abs} shows expected sensitivity in the mixing parameter, $\kappa$, of the proposed setup for absorption events of dark-photon DM for $d=5\textrm{nm}$ PbS QDs (we use QD sizes that are slightly different from those used for the DM-electron scattering results, since the available experimental data correspond to 5~nm). The left panel shows results for the coincident two-photon signal assuming the same exposures and background fluxes as those used in Fig.~\ref{fig:QDsense}. The right panel shows results for the single-photon signal under the assumption of negligible backgrounds. Existing limits from DM absorption from SENSEI~\cite{SENSEI:2020dpa}, XENON10~\cite{Bloch:2016sjj} and limits on solar emission from XENON1T~\cite{An:2020bxd} are also shown. The SENSEI~\cite{SENSEI:2020dpa} limits are rescaled by $|\epsilon|$ (the absolute value of the dielectric function) to account for in-medium effects~\cite{Mitridate:2021ctr}. Curves are dotted above $m_{A'} \ge 10~\textrm{eV}$, since for these larger masses the QD response is currently not well understood. 

Similar conclusions to the scattering case can be drawn regarding exposures and background rates and competitiveness with other experimental bounds. In particular, for the coincident two-photon case, an exposure of approximately $100$~g-year and single-photon background rates below $1$ Hz would make it possible to explore new parameter space. A more futuristic detector sensitive to single-photon events and with negligible backgrounds will be able to probe many orders of magnitude of new parameter space.

\section{Conclusions}
\label{sec:conclusions}

While there is a wide variety of proposed detection targets in the sub-GeV electron-scattering space, searches using QDs are  uniquely suited to take advantage of the scalability and optical properties of molecules, while also leveraging their tunable low-thresholds and multi-exciton dynamics of semiconductors. In this paper, we have shown that strongly confining IV-VI semiconducting nanocrystals in colloidal suspension are a promising target material, which can be deployed quickly and probe unexplored DM parameter space.

The signal consists of either a coincident two-photon event with inherently low backgrounds, or a single-photon event. For the former case, currently available photodetector technologies such as PMTs are sufficient to have sensitivity to DM that would be competitive or even supersede current direct-detection bounds. Such an experimental setup is achievable on short timescales and requires little R\&D. For the latter (single-photon) case, lower background photodetectors and dedicated background reduction techniques would be required. Such a setup is envisioned as a futuristic progression of the experiment.

For both DM scattering and bosonic DM absorption, a modest $100\,$g-year exposure of PbS QDs with a diameter of $3.3 - 5\,$nm are shown to be up to two orders of magnitude more sensitive than existing constraints for realistic background rates with the coincident two-photon setup. Smaller scale, $10\,$g-month exposures are expected to be essentially ``background-free" and present a natural characterization scale for the experiment with interesting science potential and competitive sensitivity to cutting edge detectors such as SENSEI and SuperCDMS. The single-photon setup will be able to probe many orders of magnitude of new DM parameter space for both DM scattering and for DM bosonic absorption.

Improvements over the conservative benchmarks used in this study may be achieved by lowering the steady-state background while maximizing the detection efficiency of the photodetector. This could be done by running at a deep-underground site and using intrinsically low-threshold low-background detectors such as SNSPDs, MKIDs, TESs, or skipper-CCDs. 
Such photodetectors have already been proposed for use in other direct-detection experimental setups consisting of, e.g., solid-state scintillators~\cite{Derenzo:2016fse} (such as GaAs) or molecular gases~\cite{Essig:2019kfe}. Additional possible improvements with respect to results shown in this study include characterizing the bi-exciton radiative quantum yield of the QDs via sample-averaged time-resolved photon self-correlation measurements~\cite{nair2011biexciton,beyler2014sample}. Additionally, it should be possible to experimentally optimize the multi-photon emission of the QDs by suppressing Auger-recombination or enhancing the spontaneous emission rate~\cite{beyler2014sample,taghipour2022low,dhawan2020extreme}. Yet another promising avenue is the consideration of cooperative emission effects~\cite{Arvanitaki:2017nhi} to focus the emergent photons produced after DM absorption into small surface single photon detectors.

This study is a first analysis of the physics case for a QD-based DM detector. Such a setup would utilize the scintillation properties of QDs and would benefit from a photon readout as opposed to the more conventional electron readout typical for semiconductor-based experiments. We conclude that with existing photodetector technologies and commercially available QDs (such as PbS QDs), it seems possible to deploy an intrinsically low-background experimental search for MeV-scale DM scattering events and eV-scale DM absorption events that could significantly improve on the current bounds from the leading sub-GeV DM experiments.

\section{Acknowledgements}
We thank Juan Collar, Dan Baxter, Yonatan Kahn, and Konstantin Likharev for useful discussions. The work of C.B.~was supported in part by NASA through the NASA Hubble Fellowship Program grant HST-HF2-51451.001-A awarded by the Space Telescope Science Institute, which is operated by the Association of Universities for Research in Astronomy, Inc., for NASA, under contract NAS5-26555  as well as by the European Research Council under grant 742104. R.E.~acknowledges support from DoE Grant DE-SC0009854, Simons Investigator in Physics Award 623940, and the US-Israel Binational Science Foundation Grant No.~2016153. O.S.~is supported by the DOE under Award Number DE-SC0007968 and the Binational Science Foundation (grant~No.~2018140). H.R.~acknowledges the support from the Simons Investigator Award 824870, DOE Grant DE-SC0012012, NSF Grant PHY2014215, DOE HEP QuantISED award no. 100495,
and the Gordon and Betty Moore Foundation Grant
GBMF7946. We are also grateful to the organisers of the Pollica Summer Workshop,  supported by the Regione Campania, Università degli Studi di Salerno, Universit\`a degli Studi di Napoli ``Federico II'', i dipartimenti di Fisica ``Ettore Pancini'',  and ``E R Caianiello'', and ``Istituto Nazionale di Fisica Nucleare'', which allowed us to collaborate. In addition, we thank the Simons Center for Geometry and Physics and the program ``Lighting new Lampposts for Dark Matter and Beyond the Standard Model'', which also allowed us to collaborate on this paper.

\bibliography{DM}

\newpage

\onecolumngrid
\appendix

\section{Semi-analytic Model for DM-Electron Scattering in QDs}
\label{sec:semianalmod}

This appendix provides intuition as to why the scattering rates for QDs can be well approximated by scattering rates of a bulk semiconductor. Below, we present a simple semi-analytical model for the scattering of DM with the electronic states in semiconductors. However, this calculation breaks down at very small momentum transfer, $qa \ll 1$ (where $a$ is the lattice constant), precisely in the kinematic region where QDs are expected to differ from semiconductors. We approximate the result for QDs by adjusting the semiconductor calculation at low $q$ and accounting for effects of quantum confinement close to the band-edge. With these calculations, we show that differences between the two occur in regions of parameter space that are anyway kinematically forbidden.

The main conclusion is that the effects of quantum confinement are only important for radiative de-excitation of QDs but not for their excitation rates. Therefore it is typically sufficiently accurate to use numerical calculations of the bulk material in order to calculate QD excitation rates, while size-dependent radiative de-excitation rates can then be inferred from experimental results.

We begin by following previous semi-analytic approaches to DM-electron scattering in semiconductors~\cite{Lee:2015qva,Essig:2012yx,Essig:2011nj,Essig:2015cda,Essig:2016crl}. The calculation for the total directionally-averaged differential electronic excitation rate in QDs was discussed in Sec.~\ref{subsec:Rates}. For completeness, the main equations are provided here as well. The excitation rate (Eq.~\eqref{eq:diffRate} of the main text) is 
\beq
    \frac{dR_{\rm ex}}{d\ln{E_r}} = \frac{N_t \rho_{\chi}\overline{\sigma}_e} {8m_{\chi} \mu_{e\chi}} \int_{q_{\rm min}}^{q_{\rm max}} q dq \, \eta(v_{\rm min}) |F_{\rm DM}(q)|^2  |f(q,E_r)|^2\,,
    \label{eq:AP_diffRate}
\eeq
where the QD subscript has been removed from the form factor, since here we present general arguments that hold for both bulk semiconductors and for QDs. The general equation for the form factor (Eq.~\eqref{eq:fqd} of the main text) is 
\beq
|f(q,E_r)|^2 \equiv \sum_{\rm fin} \sum_{\rm init} \int dE_b \left(\frac{d n}{d E_b}\right)_{\rm init} \delta(E_r - (E_{\rm fin} - E_b)) \left \vert \int d^{3}r \,e^{-i\mathbf{q}\cdot\mathbf{r}} \Psi_{\rm fin}^{*}(\mathbf{r})\Psi_{\rm init}(\mathbf{r}) \right \vert^2,
\label{eq:AP_fqd}
\eeq
where $(dn/dE_b)_{\rm init}$ is the density of initial states (DOS) of the system~\cite{Graham:2012su,lach2002electronic}, $E_b$ is the binding energy, 
and the initial and final wavefunctions have also been written in their general forms.

For the initial states, one can take the approximation of the tight-binding model. In this approximation, small perturbations to the wavefunction from nearest-neighbor interactions are neglected, and the valence band is modeled as a sum over atomic orbitals (analogous to an LCAO model for the case of molecules). The initial electronic states are then well described by the relevant valence atomic orbital,
\begin{align}
    \Psi_{\rm init}(\mathbf{r}) =  \sum_N e^{i \mathbf{k}\cdot\mathbf{R}_N} \mathcal{Y}_{\ell m}(\Omega) \chi_{n \ell}(r)\,,
\end{align}
where $\mathbf{k}$ is the crystal momentum and $\mathbf{R}_N$ are the locations of atoms within the crystal. The  functions $\psi^{\rm RHF}_{n \ell m}(\mathbf{r}) \equiv \mathcal{Y}_{\ell m}(\Omega)\chi_{n \ell}(r)$ are isolated Roothan-Hartree-Fock atomic orbitals, $\mathcal{Y}_{\ell m}(\Omega)$ are spherical harmonics, and $\chi_{n \ell}(r)$ are radial wavefunctions, which can be expanded as a finite sum over primitive Slater-type radial functions,
\begin{align}
    \chi_{n \ell}(r) &= \sum_j c_{j \ell n} N_{j \ell}r^{(n_{j \ell} -1)}\mathrm{exp}(-Z_{j \ell}r)\,. 
\end{align}
Here, $c_{j \ell n}$ are the expansion coefficients, $N_{j \ell}$ are normalization constants, $n_{j \ell}$ are principle quantum numbers, and $Z_{j \ell}$ are effective charges. In calculations presented here, we take Roothan-Hartree-Fock functions from tabulated numerical fits to atomic valence states~\cite{bunge1993roothaan}.

\subsection*{Approximation for Bulk Semiconductors at High Momentum Transfer}

If the imparted momentum is sufficiently large, the final electronic states can be approximated as plane waves. However, one must also account for the deformation of these final states by the presence of the charged ion at the origin. The result for the form factor, Eq.~\eqref{eq:AP_fqd}, can then be rewritten as,
\beq
|f(q,E_r)|^2 \equiv \sum_{n\ell} \int dE_b \left(\frac{d n}{d E_b}\right)_{n\ell} F_{\rm Fermi}(E_r - E_b) \left \vert f^{n \ell}(q, E_r, E_b) \right \vert^2\,,
\label{eq:AP_fqd2}
\eeq
where the Fermi function accounts for the wavefunction deformation close to the origin, and is given by 
\begin{equation}
    F_{\rm Fermi} = \frac{2\pi\nu}{1-\exp{(-2\pi \nu)}} \,\,\,\,\, ; \,\,\,\,\, \nu = \frac{\alpha m_e Z_{\rm eff}}{\sqrt{2 m_e (E_r - E_b)}}\,.
\end{equation}
 
$Z_{\rm eff}$ is the effective charge of the ion at the origin, which we conservatively take as $Z_{\rm eff}=1$ to account for outer-shell electrons~\cite{Lee:2015qva,Essig:2011nj,Essig:2012yx}.

The last term in Eq.~\eqref{eq:AP_fqd2} is related to the wavefunction overlap. For recoil energies that are of order or larger than the binding energy of the initial state, the form factor is approximately~\cite{Lee:2015qva,Essig:2012yx},
\begin{equation}
    |f^{n\ell}(q,E_r,E_b)|^{2} =\frac{\left(2\ell+1\right)2m_e (E_r - E_b)}{4\pi^{3} q}
    \int_{|\sqrt{2m_e E_r}-q|}^{\sqrt{2m_e E_r}+q} k |\tilde{\chi}_{n \ell}(k)|^2 \,dk\,,
    \label{eq:AP_fqd3}
\end{equation}
where $\tilde{\chi}_{n \ell}(k)$ is the fourier transform of $\chi_{n \ell}(r)$~\cite{belkic1989unified}. 

Note that accounting for the binding energy corresponds to energy-dependent limits of integration and acts to enhance the rate. This is a correction to the approach in Refs.~\cite{Lee:2015qva,Essig:2012yx}, which was adapted from the arguments in Ref.~\cite{Essig:2016crl}.

\subsection*{Approximation for Quantum Dots at Low Momentum Transfer}
In order to account for the effects of quantum confinement at low momentum transfer, one should correct the form factor calculation of Eq.~\eqref{eq:AP_fqd3}. Since the final states for that calculation were taken as plane waves, the basis used for the calculation is not orthogonal and the form factor does not approach zero as $q \to 0$, as it should. This can be corrected to first order in $\mathbf{q} \cdot \mathbf{r}$ by using the dipole approximation, $e^{i \mathbf{q} \cdot \mathbf{r}}\approx 1+i \mathbf{q} \cdot \mathbf{r}$, from which it follows that $\bra{\psi_{\rm fin}}e^{i \mathbf{q} \cdot  \mathbf{r}}\ket{\psi_{\rm init}} \approx i \mathbf{q} \cdot \bra{\psi_{\rm fin}}\mathbf{r}\ket{\psi_{\rm init}}$ for an orthogonal basis.

The dipole matrix element between the initial (valence) and final (conduction) bands of a semiconductor is given in $\mathbf{k}\cdot\mathbf{p}$ perturbation theory by 
\begin{align}
    |\bra{\psi_{\rm fin}} r \ket{\psi_{\rm init}}|^2 = \frac{(m_e - m_{\rm eff})}{2m_e m_{\rm eff} \Delta E_g}\,,
\end{align}
where $m_{\rm eff}$ is the curvature of the valence and conduction bands around their extrema. The result for $f^{n\ell}$ for the low momentum transfer regime is then 
\begin{equation}
    |f^{n\ell}(q,E_r=E_b=\Delta E_g)|^{2} = \frac{(m_e - m_{\rm eff})}{2m_e m_{\rm eff} \Delta E_g} q^2\,.
    \label{eq:AP_FF_lowq}
\end{equation}

From the results of Eqs.~\eqref{eq:AP_fqd3} and~\eqref{eq:AP_FF_lowq}, one can construct a form factor that is approximately correct both at high and low momentum transfer,
\begin{equation}
    |f^{n \ell}(q,E_r,E_b)|^2 = \begin{cases} 
      \text{Eq.}~\eqref{eq:AP_FF_lowq} & ,qa < 1 \\
      \text{Eq.}~\eqref{eq:AP_fqd3} & ,qa\geq1. 
    \end{cases}
    \label{eq:fnl_fin}
\end{equation}

The effects of quantum confinement cause the bandgap of the semiconductor to effectively increase. As the QD becomes smaller, the envelope piece of the wavefunction  affects the transition by effectively increasing the energy of the valence and conduction states, as can be seen from the envelope wave functions in Eq.~\ref{eq:QDpsi3}. This can be modeled to first order as in Eq.~\ref{eq:EgapQD}, where the second term is the energy of confinement of the hole and electron states. Practically, the first transition energy for QDs can be precisely measured and its dependence on the QD radius, $R$, can be experimentally extracted as in Eq.~\ref{eq:QDexpEnergy}. For Silicon QDs, this relation is given by~\cite{gesese2009estimation},
\begin{align}
    \Delta E = \Delta E_g + \left( \frac{3.73}{(R/\text{nm})^{1.37}}\right).
\end{align}

\begin{figure}
    \centering
    \includegraphics[width=0.6\columnwidth]{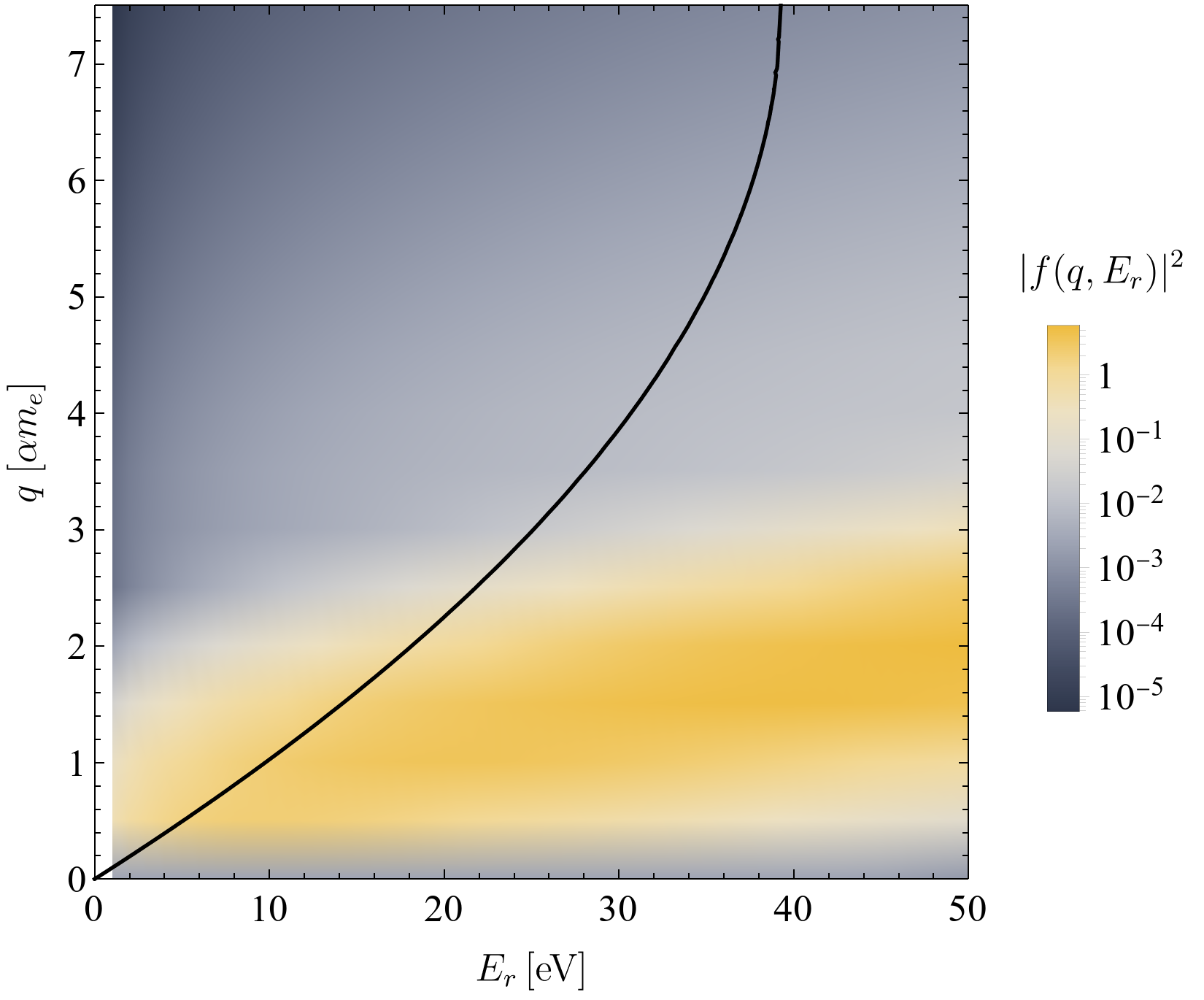}
    \caption{The Si form factor as a function of momentum transfer, $q$, and recoil energy, $E_r$, for a QD with $d=3.3$~nm accounting for the confinement via the dipole correction. The black curve delineates the minimum kinematically allowed momentum $q_{\rm min}$ for a DM mass of $10$~MeV. White regions are numerically indistinguishable from zero and fall below $\Delta E_{\text{QD}}$.}
    \label{fig:f2d}
\end{figure}

Fig.~\ref{fig:f2d} shows the form factor, $|f(q,E_r)|^2$, for Si using $|f^{n \ell}(q,E_r,E_b)|^2$ as given in Eq.~\eqref{eq:fnl_fin}. Also shown in black is the lower bound for the kinematically allowed region of the $q$-integral for a 10~MeV DM particle. Note how even for a transition energy equal to the band gap of Si, the kinematically allowed region is always significantly above the dipole part of the form factor. Indeed, rates even slightly above threshold are expected to be entirely undisturbed by the QD nature of the target.

Since the QD nature of the target has been shown not to affect the rates compared to those calculated for the bulk, we conclude that it is appropriate to use a sufficiently accurate numerical form factor for QDs as calculated for bulk semiconductors but with a modified band-gap, whose value depends on the QD radius. In other words, the number of charge carriers created in a semiconductor by a scattering event with a massive DM particle would be the same if the semiconductor is monolithic or nanoscopically disperse. 

\section{Quantum Yield}
\label{sec:QY}
 Given the single exciton quantum yield (QY$_{\gamma}$), defined as the ratio of emitted photons to primary electronic excitations, the probability of a photon exiting the bulk target (QY$_{\rm BF}$) after an excitation is given by the following,
 \begin{equation}
   \text{QY}_{\rm BF} = (1-a_{xx})\text{QY}_{\gamma}\,,
 \end{equation}
where $a_{xx}$, the probability of self absorption, can be calculated with the following expression,
\begin{equation}
    a_{x x}=\frac{\int I_{x}(\omega)\left[1-\exp \left(-\epsilon_{x}(\omega) l\right)\right] d \omega}{\int I_{x}(\omega) d \omega}\,.
\end{equation}
Here $I_{x}$ is the relative quantum intensity of the chromophore's fluorescence (i.e., photoluminescence, photoemission), $\epsilon_{x}$ is the molar extinction coefficient, and $l$ is the path length the photon must travel. Note that $I_{x}$ is distinct from the bulk fluorescence spectrum in that the bulk fluorescence spectrum is proportional to QY$_{\rm BF}$ due to self absorption. $\epsilon_{x}$ is a quantity that can be experimentally measured and theoretically predicted. From the Beer-Lambert law, the molar extinction coefficient is given by,
\begin{align}
    \left(\frac{\epsilon_{x}(\omega)}{\textrm{M}^{-1}\;\textrm{cm}^{-1}}\right)&=\frac{1}{ \ln(10)}\left(\frac{N_{A}}{\textrm{mol}^{-1}}\right)   \left(\frac{\text{mol}}{\text{M}\,\text{L}}\right) \left(\frac{\text{L}}{10^3\; \text{cm}^3}\right) \left(\frac{\sigma_{x}(\omega)}{\textrm{cm}^{2}}\right)\nonumber \\
    &=\frac{1}{ \ln(10)}\left(\frac{N_{A}}{\textrm{mol}^{-1}}\right)\left(\frac{\text{mol}}{\text{M}\,\text{L}}\right) \left(\frac{\text{L}}{10^3\; \text{cm}^3}\right) \left(\frac{\text{cm}}{\lambda_{x}(\omega)}\right) \left(\frac{\text{cm}^{-3}}{n}\right)\,,
\end{align}
where $\sigma_{x}$ is the absorption cross section for a photon of energy $\omega$, $N_{A}$ is Avogadro's number, $\lambda$ is the mean free path, and $n$ is the chromophore number density. Note that we have specified the units in order to highlight the conventions found in the physics and chemistry literature. One can think of $\epsilon_{x}$ as the inverse mean free path per unit number density. The factor of $\ln(10)$ comes from the Beer-Lambert law. The transmittance, $T$, is given as a function of absorbance, $A$, by
\begin{equation}
    T(\omega)=\frac{\phi_t(\omega)}{\phi_i(\omega)} = e^{-\sigma(\omega) n l} = 10^{-\epsilon_{x}(\omega) c l}=10^{-A(\omega)}\,,
\end{equation}
where, for a bulk sample of chromophores $x$, $\phi_i$ is the incoming radiant flux, $\phi_t$ is the transmitted radiant flux, $A(\omega)$ is defined as the negative log of the transmittance $T(\omega)$, and $c$ is the molar concentration of the chromophore. Thus, a measurement of the absorbance of a bulk sample of chromophores of known concentration is a direct measurement of the molar absorption coefficient and, in turn, its light absorption cross section. Therefore, measurements of $I_{x}(\omega)$ and $A(\omega)$ are sufficient in order to compute the probability for a signal photon to be emitted and to traverse a macroscopic sample as a result of exciting a chromophore in the target bulk.

Notice that in dilute concentrations of chromophores, after the rapid self-absorption between the emission, $I_x(\omega)$, and absorption, $\epsilon_x(\omega)$, overlap in $a_{xx}$, further self-absorption is negligible for pathlengths within about an order of magnitude around the $l$ for which the measurements where taken. The relative offset of the peak absorption and emission characteristics for a chromophore is called the Stokes' shift and is correlated with increasing QY$_{\rm BF}$~\cite{semonin2010absolute}, since this causes the overlap in $a_{xx}$ to decrease rapidly. For PbS QDs of $d=3.3$~nm, the concentration at which further self-absorption becomes important due to other factors such as quenching or aggregation is around 10  $\mu$M/L~\cite{Greben2015}.

\section{Si and PbS form factor calculations}

All results presented in this study were obtained using single electron crystal wave function orbitals and energies obtained using {\sc Quantum Espresso}~\cite{giannozzi2009quantum}.
They are obtained from density functional theory (DFT) within the local density approximation (LDA) using the Perdew-Zunger~\cite{perdew1981self} parameterization for the exchange and correlation functional. Norm conserving pesudopotentials are used for the core electronic levels
and plane wave cutoffs of up to 35~Ry (for PbS) and 30~Ry (for Si) were employed in the calculation of the Kohn-Sham orbitals.
For Si (Ge), a band gap corrected to 1.1 (0.67)~eV is applied, while for PbS we correct the band gap to be 0.42~eV. Finally, the band gap of PbS is adjusted to the appropriate value given by Eq.~\ref{eq:QDexpEnergy} for the QD calculations. 
Input files for the electronic structure calculations and $\texttt{QE-dark}$ are provided in~\cite{QEdata1,QEdata2}.

\end{document}